\documentclass[aps,prb,twocolumn,amsmath,amssymb,10pt,aps,longbibliography,superscriptaddress,citeautoscript,bibnotes,floatfix,nobibnotes]{revtex4-2}

\usepackage[utf8]{inputenc}
\usepackage[T1]{fontenc}
\usepackage{textcomp} %% provides additional text symbols

\usepackage{amsmath}   %% basic math formatting
\usepackage{amssymb}   %% this does nothing without amsfonts?
\usepackage{physics}   %% various physics notation, e.g. bra-ket
\usepackage{bm}        %% bold math symbols, e.g. greek letters which do not have a \mathbf
\usepackage{mathrsfs}  %% \mathsf font
\usepackage{mathtools} %% additional features to amsmath, e.g. dcases environment
\usepackage{caption}  %% additional control of the captions in floating environments
\usepackage{graphicx} %% additional control of \includegraphics, extension of {graphics}
\graphicspath{ {./Figures/} }

\usepackage{subfigure} %% deprecated, should be replaced with {subcaption}

\usepackage{tabularx} %% extension of {tabular} with adjustable-width columns
\usepackage{booktabs} %% internal optimizations and additional control of table environments
\usepackage{multirow} %% create tabular cells spanning multiple rows
\usepackage{dcolumn}  %% align table columns on decimal point

\usepackage[colorlinks=true,linkcolor=blue,citecolor=blue]{hyperref} 

\usepackage[normalem]{ulem} %% provides underline command, \ul
\usepackage{soul} %% spacing out (letterspacing), underlining, striking out, etc.
\usepackage{amsthm}  %% AMS formatting of {theorem} environments
\usepackage{color,xcolor} %% foreground (text, rules, etc.) and background colour management
\usepackage{lipsum}

\usepackage{comment}

\usepackage{soul}
\newtheorem{lemma}{Lemma}
\newtheorem{theorem}{Theorem}
\newtheorem{theoremAPPENDIX}{Theorem}
\newtheorem{corollary}{Corollary}
\newtheorem{corollaryAPPENDIX}{Corollary}
\newtheorem{definition}{Definition}

\begin{document}
\title{Quaternionic Hermitian Band Geometry in Four Dimensions: Realization on $S^4$ and Obstruction on $T^4$}

\author{Dongju Hwang}
\thanks{These authors contributed equally.}
\affiliation{Department of Physics, Korea Advanced Institute of Science and Technology, Daejeon 34141, Republic of Korea}

\author{Minjae Yu}
\thanks{These authors contributed equally.}
\affiliation{Department of Physics, Korea Advanced Institute of Science and Technology, Daejeon 34141, Republic of Korea}

\author{Gil Young Cho}
\thanks{gilyoungcho@kaist.ac.kr}
\affiliation{Department of Physics, Korea Advanced Institute of Science and Technology, Daejeon 34141, Republic of Korea}
\affiliation{Center for Artificial Low Dimensional Electronic Systems, Institute for Basic Science, Pohang 37673, Korea}

\begin{abstract}
We establish a realization-obstruction dichotomy for quaternionic Hermitian band geometry in four-dimensional parameter spaces: the minimal-charge lowest Landau level on $S^{4}$ provides a global realization, whereas an everywhere nondegenerate saturated realization induced by a single occupied quaternionic band is obstructed on $T^{4}$. An antiunitary symmetry $\mathcal{J}$ satisfying $\mathcal{J}^{2}=-\bm{1}$ makes the occupied doublet a quaternionic band. Within this setting, we formulate the quaternionic Wirtinger inequality as a band-geometric bound involving the quantum metric and the second Chern density. At every nondegenerate saturation point, the canonical geometry of the quaternionic projective space pulls back to a compatible quaternionic structure on the parameter space; if these conditions hold everywhere, the parameter space acquires quaternionic Hermitian band geometry.
On $S^{4}$, we express the minimal-charge states as quaternionic Perelomov coherent states and establish everywhere nondegenerate saturation, thereby realizing quaternionic Hermitian band geometry. On $T^{4}$, by contrast, a minimal four-band system saturates the inequality everywhere, but topology forces the quantum metric to become degenerate somewhere, obstructing a globally induced quaternionic structure. An explicit lattice Dirac Hamiltonian exhibits this obstruction. Adding unoccupied bands cannot remove this obstruction when the inequality is saturated everywhere, since the image of any nondegenerate saturated projector remains confined to a fixed $\mathbb{H}P^{1}$. These results provide a symmetry-aware framework for non-Abelian band geometry and show how parameter-space topology constrains the global realization of quaternionic Hermitian band geometry.
\end{abstract}

\maketitle

%\tableofcontents

\section{Introduction} \label{section 1}
Quantum geometry has emerged as a central framework for characterizing the information encoded in quantum states beyond their energy spectra and topological invariants. The quantum geometric tensor unifies the quantum metric, which measures the distance between nearby quantum states, and the Berry curvature, which describes their geometric phase. These quantities have been connected to a broad range of physical phenomena, including nonlinear response~\cite{Nonlinear1,Nonlinear2,optical1,bzdusek2025,Jankowski2025PRR}, many-body quantum phenomena~\cite{Superfluidity1,Exciton1,Jankowski2025Natcomm}, and fractional topological phases~\cite{Fractional1, Fractional2}. Recent experiments have further demonstrated direct access to quantum geometric quantities in solid-state systems~\cite{Pump2D1,Pump2D2,SunJeKim2025DirectMeasurement}. Quantum geometry should therefore be regarded not merely as an abstract characterization of wavefunctions, but as experimentally accessible geometric information encoded in quantum matter.

A particularly fruitful development in two dimensions is the identification of K\"ahler geometry through the compatibility between the quantum metric and Berry curvature \cite{Mera2021KahlerBandLL}. An isolated band over a two-dimensional parameter space $X$ defines a quantum state map $P:X \rightarrow \mathbb{C}P^{N-1}$. 
The Fubini-Study metric and symplectic form induce the quantum metric $g= P^{*}g_{\mathrm{FS}}$ and Berry curvature $F= P^{*}\mathcal{F}$ on $X$ \cite{Provost1980Riemannian,Simon1983HolonomyBerryPhase}:
\begin{align}
& g  = \mathrm{Tr}(P\mathrm{d}P\mathrm{d}P), \quad
F 
= i\mathrm{Tr}(P\mathrm{d}P \wedge \mathrm{d}P).
\end{align}
These tensors satisfy the Wirtinger inequality, whose nondegenerate saturation implies the existence of a complex structure $J$ on $X$ for which $P$ is compatible with the canonical complex structure $J_{\mathrm{FS}}$ on $\mathbb{C}P^{N-1}$:
\begin{align}
& \mathrm{d}P\circ J = J_{\mathrm{FS}} \circ \mathrm{d}P.
\end{align}
This compatibility places lowest Landau level (LLL) states and K\"ahler bands within a common geometric framework, thereby directly relating quantum geometry, complex geometry, and band topology~\cite{Mera2021KahlerBandLL,Liu2025KahlerBandLL}, and has motivated systematic constructions of ideal Chern and geometrically flat K\"ahler bands~\cite{Mera2024bKahlerBandLL,FlatBand1,FlatBand2,FlatBand3,Mera2024aKahlerBandLL}.

We investigate a quaternionic extension of two-dimensional K\"ahler band geometry to four-dimensional parameter spaces. Specifically, we ask when the quantum metric and non-Abelian Berry curvature of a quaternionic band induce a compatible geometry on the underlying parameter space. A complex rank-two occupied subspace preserved by an antiunitary symmetry $\mathcal{J}$ satisfying $\mathcal{J}^{2}=-\bm{1}$ can be regarded as a quaternionic line and carries an $\mathrm{SU}(2)$ Berry connection \cite{JandSU21,JandSU22,Bellucci2012IsospinHP}. We refer to the corresponding family of occupied doublets over the parameter space as a quaternionic band. In Bloch systems, the required antiunitary structure may be realized by parity-time $(\mathcal{PT})$ symmetry that preserves the crystal momentum and satisfies $(\mathcal{PT})^{2}=-\bm{1}$ \cite{Alexander_2024_WilsonLoopNonAbelianBerry}. A canonical continuum setting is provided by the Yang monopole and the four-dimensional quantum Hall effect, where the second Hopf fibration identifies $S^{4} \cong \mathbb{H}P^{1}$ as the natural parameter space of quaternionic quantum states \cite{Yang1978GeneralizationDiracMonopoleToSU2,Zhang2001FourQHE,Hasebe2010Hopf}. Although these structures establish the quaternionic character of the occupied fibers, they do not by themselves determine a compatible geometry on the parameter space. Guided by the metric-curvature compatibility underlying K\"ahler band geometry in two dimensions, we formulate the corresponding four-dimensional constraint as a quaternionic analogue of the Wirtinger inequality.

Our main results are as follows. First, building on the four-dimensional LLL construction and its quaternionic coherent-state formulation~\cite{Zhang2001FourQHE,Hasebe2010Hopf,Hasebe2020YangMonopoleAndBPSTInstanton,Perelomov2002CoherentState,Perelomov1986BookCoherentStates,Kuratsuji1990QuaternionicPerelomovCoherentState}, we recast the minimal-charge LLL states on $S^{4}$ as quaternionic Perelomov coherent states. This construction identifies the corresponding parameter space directly with the quaternionic projective line $\mathbb{H}P^{1}\cong S^{4}$, thereby placing the $S^{4}$ LLL within the same geometric framework used to describe quaternionic bands. 
Second, we employ the quaternionic analogue of the Wirtinger inequality~\cite{Tasaki1985WirtingerInequality,Tasaki1986WirtingerIneqaulity}, which we henceforth refer to as the quaternionic Wirtinger inequality. 
We show that, at each point where the induced quantum metric is nondegenerate, saturation determines a compatible linear quaternionic structure on the tangent space. When the inequality is saturated everywhere and the induced metric remains nondegenerate throughout the parameter space, the metric and the resulting quaternionic structure define what we call quaternionic Hermitian band geometry. For the minimal-charge $S^{4}$ LLL, these conditions hold everywhere, thereby realizing quaternionic Hermitian band geometry; in fact, the induced geometry is the canonical quaternionic K\"ahler geometry of $\mathbb{H}P^{1}$. 
Third, we turn to the global realizability of this band-geometric structure for Bloch states over the four-dimensional Brillouin zone (BZ), topologically $T^{4}$. For a minimal four-band Bloch system, we show that this inequality is saturated at every momentum, while the quantum metric must become degenerate somewhere in the BZ. We demonstrate this obstruction explicitly in a four-dimensional lattice Dirac model~\cite{QHZ2008LatticeDiracModel}, in which the loss of metric rank prevents the corresponding projector map from remaining immersive throughout the BZ.
Finally, using the local $\mathbb{H}P^{1}$ rigidity~\cite{Yang2026QKgeometryTRS}, we show that enlarging the Hilbert space by adding additional unoccupied bands does not remove this obstruction for a single occupied quaternionic band. These results therefore distinguish the local compatibility implied by saturation from the global realizability of quaternionic Hermitian band geometry.

Because the $S^{4}$ result invokes quaternionic K\"ahler geometry in real dimension four, this terminology requires particular care. For a quaternionic Hermitian manifold $X$ of real dimension $4n$ with $n \ge 2$, denote its metric by $g$ and its fundamental 4-form by $\Omega$. The quaternionic K\"ahler condition may then be expressed as $\nabla^{g}\Omega=0$, where $\nabla^{g}$ is the Levi-Civita connection of $g$. 
For $n=1$, however, this condition becomes automatic because $\Omega$ is the Riemannian volume form, just as the K\"ahler condition is automatic in real dimension two. 
Consequently, some authors retain this formally valid but nonrestrictive extension \cite{Ivanov2002AlmostIsQuaternioinic,Gray1969Note}, while others require the metric to be Einstein and self-dual to obtain a nontrivial four-dimensional notion \cite{Swann1990QK,besse2007EinsteinManifolds,Haydys2008}. 
Following the four-dimensional curvature criterion discussed by Swann \cite{Swann1990QK}, we adopt the latter convention, under which quaternionic K\"ahler geometry is distinguished from quaternionic Hermitian geometry by these additional curvature requirements. Further details are provided in Appendix \ref{appendix A}.

The remainder of this paper is organized as follows. In Sec. \ref{section 2}, we develop the quaternionic coherent-state construction on $S^4$. In Sec. \ref{section 3}, we formulate the quaternionic Wirtinger inequality, relate its saturation to the induced quantum geometry, and establish the resulting quaternionic K\"ahler structure on $S^{4}$. In Sec. \ref{section 4}, we establish the global obstruction on $T^4$ and its extension to an arbitrary number of ambient bands. Finally, in Sec. \ref{section 5}, we summarize our results. 
Appendix \ref{appendix A} reviews basic quaternionic linear algebra and the geometric definitions, Appendix \ref{appendix B} collects auxiliary lemmas and proofs, and Appendices \ref{appendix C}--\ref{appendix E} provide technical details of the coherent-state and model calculations.

\section{Quaternionic Coherent-State Realization on the Four-Sphere}
\label{section 2}
In Perelomov's construction, the Lie group $G$ acts linearly and unitarily on the quaternionic Hilbert space containing the reference state $|\psi_{e}\rangle$ \cite{Perelomov2002CoherentState,Perelomov1986BookCoherentStates}. On the four-sphere $S^{4}$, the corresponding Lie group is $G=\mathrm{Sp}(2)$, the coherent-state orbit is represented by
\begin{align}
& |\psi_{g}\rangle = T(g)|\psi_{e}\rangle \in \mathbb{H}^{2}
\end{align}
for $g \in \mathrm{Sp}(2)$ \cite{Kuratsuji1990QuaternionicPerelomovCoherentState}. Here, $T$ is an irreducible quaternionic-unitary representation. For the gauge group $\Lambda = \mathrm{Sp}(1) \cong \mathrm{SU}(2)$, the stabilizer
\begin{align}
& H = \{ h \in G ; T(h)|\psi_{e}\rangle = |\psi_{e}\rangle q \text{ for some } q\in \Lambda\}
\end{align}
is found to be $H=\mathrm{Sp}(1) \times \mathrm{Sp}(1)$. Hence, the parameter space is
\begin{align}
& X= G/H = \mathrm{Sp}(2)/\mathrm{Sp}(1)\times\mathrm{Sp}(1) \cong S^{4} \cong \mathbb{H}P^{1}. 
\end{align}
For the two-sphere $S^{2}$, Haldane considered $G=\mathrm{SU}(2)$, $\Lambda=H=\mathrm{U}(1)$, giving the parameter space \cite{Haldane1983QHE}:
\begin{align}
& X= G/H = \mathrm{SU}(2)/\mathrm{U}(1) \cong S^{2} \cong \mathbb{C}P^{1}. 
\end{align}

In the $S^{4}$ setting, some previous studies used $\mathbb{C}P^{3}$ to parameterize the complex rays in $\mathbb{C}^{4} \cong \mathbb{H}^{2}$ retaining an additional $\mathrm{Sp}(1)/\mathrm{U}(1) \cong S^{2}$ coordinate specifying a complex direction within a quaternionic line \cite{Karabali2002QHECP1CP2,Fabinger2002CP3String,Bellucci2003FourDimCp3}. By contrast, $\mathbb{H}P^{1}$ parameterizes the quaternionic lines themselves and quotients out this internal $\mathrm{Sp}(1)$ freedom:
\begin{align}
& \mathbb{C}P^{3} = \frac{\mathrm{Sp}(2)}{\mathrm{Sp}(1)\times\mathrm{U}(1)}, \quad 
\mathbb{H}P^{1} = \frac{\mathrm{Sp}(2)}{\mathrm{Sp}(1)\times\mathrm{Sp}(1)}.
\end{align}
These descriptions are complementary rather than contradictory. The $\mathbb{C}P^{3}$ formulation is appropriate when the internal isospin orientation is retained explicitly, whereas the $\mathbb{H}P^{1}$ formulation directly describes the complex rank-two quantum subspace and its $\mathrm{SU}(2)$ Berry geometry. For the purpose of describing a quaternionic band, the latter description is therefore more natural. The general quaternionic coherent-state construction and its complex-frame realization, including the stabilizer calculation and the explicit coset representative used below, are detailed in Appendix \ref{appendix C}.

To parameterize the minimal Yang-monopole sector with $\mathrm{SU}(2)$ isospin $I=1/2$, the quaternionic reference state $|\psi_{e}\rangle = (1,0)^{T} \in \mathbb{H}^{2}$ is represented by the corresponding complex frame \cite{Kuratsuji1990QuaternionicPerelomovCoherentState}
\begin{align}
& \Psi_{e} = \begin{pmatrix} \bm{1}_{2} \\ 0 \end{pmatrix} \in \mathbb{C}^{4\times2}.
\end{align}
Let $\mathrm{SP} \in S^{4}$ denote the south pole. For each $x\in S^{4}\backslash\{\mathrm{SP}\}$, we choose a local coset representative $g_{x} \in \mathrm{Sp}(2)$ that maps the north pole to $x$, and the quaternionic coherent state $|\psi_{x}\rangle = T(g_{x})|\psi_{e}\rangle$ is represented by the local complex frame \cite{Zhang2001FourQHE, Hasebe2020YangMonopoleAndBPSTInstanton}
\begin{align}
& \Psi_{x} = T(g_{x})\Psi_{e}
= \begin{pmatrix} 
\sqrt{\frac{1+x_{5}}{2}}\bm{1}_{2} \\
\sqrt{\frac{1}{2(1+x_{5})}}(x_{4}\bm{1}_{2}-ix_{i}\sigma_{i})
\end{pmatrix}.
\end{align}
The complex realization above makes manifest the fixed antiunitary structure underlying the quaternionic representation. Specifically, we define the operator
\begin{align}
& \mathcal{J}=-(\bm{1}_{2}\otimes i\sigma_{2})\mathcal{K},
\end{align}
where $\sigma_{i}$ are the Pauli matrices and $\mathcal{K}$ denotes the complex conjugation operator. This operator satisfies 
\begin{align} \label{commutation of J and T}
& \mathcal{J}^{2}=-\bm{1}_{4}, \quad \mathcal{J}T(g)\mathcal{J}^{-1}=T(g),
\end{align}
for all $g \in \mathrm{Sp}(2)$. Writing the two columns of the coherent-state frame as
\begin{align}
& \Psi_{x} = \Bigr[ |\psi_{x}^{(1)}\rangle,\ |\psi_{x}^{(2)}\rangle \Bigr] \in \mathbb{C}^{4\times2},
\end{align}
the identity $\Psi_{x}^{\dagger}\Psi_{x}=\bm{1}_{2}$ shows that they are orthonormal, while the antiunitary structure gives
\begin{align} \label{J changes sign}
& \mathcal{J}|\psi_{x}^{(1)}\rangle = |\psi_{x}^{(2)}\rangle, \quad
\mathcal{J}|\psi_{x}^{(2)}\rangle = -|\psi_{x}^{(1)}\rangle.
\end{align}
Hence, the columns of $\Psi_{x}$ span a $\mathcal{J}$-invariant complex two-plane. Although $\Psi_{x}$ is only a local frame, its $\mathrm{SU}(2) \cong \mathrm{Sp}(1)$ gauge freedom drops out of the projector. We therefore obtain the globally defined, gauge-invariant projector
\begin{align} 
& P(x) =\Psi_{x}\Psi_{x}^{\dagger}
= |\psi_{x}^{(1)}\rangle\langle\psi_{x}^{(1)}| + |\psi_{x}^{(2)}\rangle\langle\psi_{x}^{(2)}|.
\end{align}
Since $\mathcal{J}$ exchanges the two columns up to a sign according to Eq. (\ref{J changes sign}), this projector satisfies $\mathcal{J}P(x)\mathcal{J}^{-1}=P(x)$.

The projector-level $\mathcal{J}$-invariance exhibited by the $S^{4}$ coherent-state projector admits a direct extension to general Bloch bands. Accordingly, in a general $2N$-dimensional complex Bloch system, we consider a smooth rank-two projector $P:X \rightarrow \mathrm{Gr}_{2,2N}^{\mathbb{C}}$ for which there exists a fixed antiunitary operator $\mathcal{J}$ satisfying 
\begin{align} \label{commutation of J and P in T4}
& \mathcal{J}^{2}=-\bm{1}_{2N}, \quad\mathcal{J}P(x)\mathcal{J}^{-1} = P(x),
\end{align}
for all $x \in X$. This condition is precisely the complex formulation of a single occupied quaternionic band. To make this correspondence explicit, we introduce the map 
\begin{align}
& \eta: \mathrm{Gr}_{1,N}^{\mathbb{H}} \rightarrow (\mathrm{Gr}_{2,2N}^{\mathbb{C}})^{\mathcal{J}},
\end{align}
which regards a quaternionic line in $\mathbb{H}^{N}$ as a $\mathcal{J}$-invariant complex two-plane in $\mathbb{C}^{2N}$. By Lemma \ref{lemma 1}, $\eta$ is a diffeomorphism onto the $\mathcal{J}$-invariant locus. Moreover, Lemma \ref{lemma 2} shows that Eq. (\ref{commutation of J and P in T4}) holds precisely when $P$ arises from a quaternionic rank-one projector $P_{\mathbb{H}}: X \rightarrow \mathbb{H}P^{N-1}$, in which case $P = \eta \circ P_{\mathbb{H}}$.

\section{Quaternionic Wirtinger Inequality in Band Geometry}
\label{section 3}
Before formulating the quaternionic inequality, we briefly recall its complex two-dimensional counterpart. Let $X$ be an oriented two-dimensional parameter space, and let $P:X\rightarrow \mathbb{C}P^{N-1}$ be the complex rank-one projector map. We define the pullback metric and 2-form by $g = P^{*}g_{\mathrm{FS}}$, $\omega = P^{*}\omega_{\mathrm{FS}}$. At each immersion point $x\in X$, the Wirtinger inequality gives \cite{Harvey1982CalibratedGeometry}
\begin{align}
& |\omega| \le \mathrm{vol}_{g},
\end{align}
where $\mathrm{vol}_{g}$ denotes the Riemannian volume form induced by $g$. Here, the pullback form $\omega=P^{*}\omega_{\mathrm{FS}}$ should be distinguished from the intrinsic fundamental form determined by the pullback metric $g$ and a complex structure $J$ on $X$,
\begin{align}
& \widetilde{\omega}(u,v) = g(Ju,v) \text{ for } u,v\in T_{x}X.
\end{align}
At every point where $P$ is an immersion, the inequality is saturated if and only if these two forms agree up to orientation $|\widetilde{\omega}| = |\omega|$, or equivalently, the holomorphic condition $\mathrm{d}P \circ J = J_{\mathrm{FS}} \circ \mathrm{d}P$ is satisfied. Finally, using the convention $F=-2\omega$, the inequality takes the familiar band-geometric form 
\begin{align} \label{2D Wirtinger inequality}
& \frac{1}{2}|F_{12}(x)| \le \sqrt{\det g(x)}.
\end{align}
Here, $F_{12}(x)$ denotes the coefficient of $\mathrm{d}x^{1}\wedge\mathrm{d}x^{2}$ in the Berry curvature $F$ with respect to any positively oriented local coordinates $(x^{1},x^{2})$ around $x$. When $P$ is an immersion and the inequality is saturated everywhere, the induced geometry is the familiar K\"ahler band geometry on $X$ \cite{Mera2021KahlerBandLL,Mera2024bKahlerBandLL,Liu2025KahlerBandLL}.

We now pass from the complex to the quaternionic setting. Although the individual fundamental 2-forms $\omega_{\mathrm{FS}}^{i}$ depend on the choice of a local admissible frame, their $\mathrm{SO}(3)$-invariant combination defines a globally defined Kraines form \cite{Kraines1966TopologyOfQuaternionicManifolds}. With $T_{i}=-i\sigma_{i}/2$, the universal $\mathrm{SU}(2)$ Berry curvature satisfies $\mathcal{F}^{i}=-2\omega^{i}_{\mathrm{FS}}$, as shown in Lemma \ref{lemma 3}. Consequently, the Kraines form can be written equivalently as 
\begin{align}
& \Omega_{\mathrm{FS}} 
= \frac{1}{6}\sum_{i=1}^{3} \omega^{i}_{\mathrm{FS}} \wedge \omega^{i}_{\mathrm{FS}}
= -\frac{1}{12}\mathrm{Tr}(\mathcal{F} \wedge \mathcal{F}).
\end{align}
The quaternionic Wirtinger inequality for $\Omega_{\mathrm{FS}}$ therefore takes the following form in terms of the Berry curvature \cite{Tasaki1985WirtingerInequality,Tasaki1986WirtingerIneqaulity}.

\begin{theorem} \label{theorem 1}
Let $X$ be an oriented four-manifold and suppose that $P$ factors through a quaternionic projector map $P_{\mathbb{H}}: X \rightarrow \mathbb{H}P^{N-1}$. Let $g=P_{\mathbb{H}}^{*}g_{\mathrm{FS}}$ be the pullback Fubini-Study metric, and let $F$ be the associated $\mathrm{SU}(2)$ Berry curvature. Then, for every $x \in X$, 
\begin{align} \label{quaternionic Wirtinger inequality}
& \frac{1}{12}|[\mathrm{Tr}(F\wedge F)]_{1234}(x)| \le \sqrt{\det g(x)}.
\end{align}
At each point $x\in X$, the following two cases occur. \\
\noindent
$\mathrm{(i)}$ $\mathrm{rank}(\mathrm{d}P_{\mathbb{H};x}) < 4$: both sides vanish, and hence the equality holds automatically. \\
\noindent
$\mathrm{(ii)}$ $\mathrm{rank}(\mathrm{d}P_{\mathbb{H};x}) = 4$: the equality holds if and only if the four-dimensional subspace $\mathrm{d}P_{\mathbb{H};x}(T_{x}X)$ is invariant under the canonical quaternionic structure $\mathcal{Q}_{\mathrm{FS}} \subset \mathrm{End}(T\mathbb{H}P^{N-1})$. Equivalently, the rank-three subspace
\begin{align} \label{quaternionic holomprhic}
& \mathcal{Q}_{x} := (\mathrm{d}P_{\mathbb{H};x})^{-1} \circ \mathcal{Q}_{\mathrm{FS}} \circ \mathrm{d}P_{\mathbb{H};x} \subset \mathrm{End}(T_{x}X)
\end{align}
determines a well-defined $g_{x}$-compatible linear quaternionic structure on $T_{x}X$. 
\end{theorem}

Here, $[\mathrm{Tr}(F \wedge F)]_{1234}(x)$ denotes the coefficient of $\mathrm{d}x^{1}\wedge\mathrm{d}x^{2}\wedge\mathrm{d}x^{3}\wedge\mathrm{d}x^{4}$ in $\mathrm{Tr}(F \wedge F)$ with respect to any positively oriented local coordinates $(x^{1},x^{2},x^{3},x^{4})$ around $x$. The geometric content of Theorem \ref{theorem 1} is pointwise and differs between degenerate and nondegenerate points. At a degenerate point, both sides of Eq. (\ref{quaternionic Wirtinger inequality}) vanish, so its saturation is automatic and provides no quaternionic information. At a nondegenerate point, by contrast, saturation identifies the tangent space of $X$ with a quaternionic subspace. In this sense, Eq. (\ref{quaternionic holomprhic}) plays the same role as the holomorphic condition $\mathrm{d}P \circ J = J_{\mathrm{FS}} \circ \mathrm{d}P$ in K\"ahler geometry, with the single complex structure $J$ replaced by the rank-three quaternionic structure $\mathcal{Q}$. If saturation holds everywhere and the pullback metric remains nondegenerate throughout $X$, these pointwise structures assemble into a global quaternionic structure.

\begin{corollary} \label{corollary 1}
Let the assumptions of Theorem \ref{theorem 1} hold. Suppose that the quaternionic Wirtinger inequality is everywhere saturated and that $\det g(x) > 0$ for every $x \in X$. Then, $P_{\mathbb{H}}$ is an immersion, and 
\begin{align}
& \mathcal{Q} = \bigsqcup_{x\in X} \mathcal{Q}_{x} \subset \mathrm{End}(TX)
\end{align}
defines a smooth rank-three quaternionic structure on $X$ compatible with the metric $g$. Consequently, $(X,g,\mathcal{Q})$ is a quaternionic Hermitian manifold.
\end{corollary}

We refer to the geometry $(X,g,\mathcal{Q})$ obtained under the hypotheses of Corollary \ref{corollary 1} as quaternionic Hermitian band geometry induced by $P_{\mathbb{H}}$. In four dimensions, an oriented Riemannian metric already determines a compatible rank-three bundle. The nontrivial content of Corollary \ref{corollary 1} is therefore band-geometric: nondegenerate saturation identifies this structure with that induced from the canonical target structure $\mathcal{Q}_{\mathrm{FS}}$ through $\mathrm{d}P_{\mathbb{H}}$. Under the four-dimensional convention adopted in Definition \ref{definition 7} of Appendix \ref{appendix A}, this conclusion does not itself imply quaternionic K\"ahler geometry; the induced metric must additionally be Einstein and self-dual. 

The pointwise inequality in Eq. (\ref{quaternionic Wirtinger inequality})  requires neither saturation nor nondegeneracy. At metric-degenerate points, both sides vanish, so such points do not invalidate the inequality. Integrating it over a closed four-manifold gives the following global bound.

\begin{corollary} \label{corollary 2}
Let the assumptions of Theorem \ref{theorem 1} hold. If $X$ is closed, then
\begin{align} \label{lower bound}
& \frac{2\pi^{2}}{3}|\mathcal{C}_{2}| \le \int_{X}\sqrt{\det g(x)}\ \mathrm{d}^{4}x,
\end{align}
where 
\begin{align}
& \mathcal{C}_{2} = -\frac{1}{8\pi^{2}} \int_{X} \mathrm{Tr}(F\wedge F) \in \mathbb{Z}
\end{align}
is the second Chern number of the occupied complex rank-two bundle defined by $P$.
Equality holds if and only if the quaternionic Wirtinger inequality is saturated at every point of $X$ and $\mathrm{Tr}(F\wedge F)$ does not change sign with respect to the orientation of $X$.
\end{corollary}

If $X$ is closed and connected, the nondegenerate saturation assumed in Corollary \ref{corollary 1} implies that $\mathrm{Tr}(F\wedge F)$ is nowhere zero and hence has a fixed sign with respect to the orientation of $X$. Therefore, the global bound in Corollary \ref{corollary 2} is also saturated. The converse need not hold: the equality in Corollary \ref{corollary 2} requires everywhere saturation and a fixed sign of $\mathrm{Tr}(F \wedge F)$, but it still permits metric-degenerate points. Thus, saturation of the global bound alone does not imply that $P_{\mathbb{H}}$ induces nondegenerate quaternionic Hermitian band geometry on $X$.

The minimal-charge coherent-state construction on $S^{4}$ realizes the stronger, nondegenerate form of saturation. We now return to the coherent-state projector $P_{\mathbb{H}}:S^{4}\rightarrow\mathbb{H}P^{1}$ constructed in Sec. \ref{section 2}. Appendix \ref{appendix D} verifies that the complex-projector calculation using $P=\eta\circ P_{\mathbb{H}}$ reproduces the metric and Berry curvature pulled back through $P_{\mathbb{H}}$. In the stereographic coordinates $y_{\mu}$ on $S^{4}\backslash\{\mathrm{SP}\}$, these are 
\begin{align} \label{metric of sphere}
& g_{\mu\nu} = \frac{2\delta_{\mu\nu}}{(1+y^{2})^{2}}, \quad
F^{i}_{\mu\nu} = -\frac{4\eta^{i}_{\mu\nu}}{(1+y^{2})^{2}},
\end{align}
where $\eta^{i}_{\mu\nu}$ are the self-dual 't Hooft symbols. These expressions saturate the quaternionic Wirtinger inequality:
\begin{align} \label{saturation of sphere}
& \frac{1}{12}|[\mathrm{Tr}(F\wedge F)]_{1234}| = \sqrt{\det g}
= \frac{4}{(1+y^{2})^{4}}>0.
\end{align}
The metric in Eq. (\ref{metric of sphere}) is one half of the standard round metric on $S^{4}$ and therefore extends smoothly and nondegenerately over the entire four-sphere. Together with the pointwise saturation in Eq. (\ref{saturation of sphere}), Corollary \ref{corollary 1} shows that $P_{\mathbb{H}}$ induces quaternionic Hermitian band geometry on $S^{4}$. Since the Einstein condition and self-duality of the round metric are preserved under constant rescaling \cite{besse2007EinsteinManifolds}, $g$ is also Einstein and self-dual. Hence, the induced geometry is quaternionic K\"ahler. The broader band-geometric significance of the additional curvature conditions remains to be understood. Appendix \ref{appendix D} further shows that the second Chern density has a fixed sign. Since $S^{4}$ is closed, the everywhere nondegenerate saturation established above therefore satisfies the equality condition of Corollary \ref{corollary 2}. Using $\mathcal{C}_{2}=1$, we obtain
\begin{align}
& \int_{S^{4}}\sqrt{\det g}\ \mathrm{d}^{4}x
= \int_{\mathbb{R}^{4}} \frac{4\mathrm{d}^{4}y}{(1+y^{2})^{4}} = \frac{2\pi^{2}}{3}|\mathcal{C}_{2}|.
\end{align}

The calculations above concern the minimal-charge sector $I=1/2$. On $S^{2}$, higher-charge coherent states are obtained by symmetric tensor product powers of the minimal spinor; their quantum metric and Berry curvature are rescaled by the same overall factor, so the saturation of Eq. (\ref{2D Wirtinger inequality}) is preserved.
On $S^{4}$, by contrast, the minimal multiplet has complex dimension two, or quaternionic dimension one, allowing the projector to be identified with a map into $\mathbb{H}P^{1} \cong S^{4}$. For $I>1/2$, the multiplet has complex dimension $2I+1>2$ and is no longer a quaternionic line, so the target must be replaced by an appropriate higher-rank Grassmannian and the present $\mathbb{H}P^{1}$ identities do not extend directly.

\section{Global Obstruction on the Four-Torus}
\label{section 4}
The preceding construction shows that the minimal-charge LLL projector induces quaternionic Hermitian band geometry on $S^{4}$. We now ask whether an occupied Bloch-band projector can likewise induce quaternionic Hermitian band geometry on $T^{4}$. Before considering systems with an arbitrary number of bands, we begin with the minimal four-band model at half-filling. Its two occupied complex bands form a single quaternionic band. The corresponding complex projector map takes values in the $\mathcal{J}$-invariant locus and factors uniquely as
\begin{align}
& P_{\mathbb{H}}: T^{4} \rightarrow \mathbb{H}P^{1}\cong S^{4}.
\end{align}
This factorization leads to a quaternionic analogue of the obstruction encountered in a two-dimensional two-band system, where the determinant of the quantum metric must vanish somewhere on $T^{2}$ \cite{Mera2021KahlerBandLL}.

\begin{theorem} \label{theorem 2}
Consider a gapped four-dimensional four-band Bloch Hamiltonian $h(\bm{\mathrm{k}})$ at half-filling with a fixed antiunitary operator $\mathcal{J}:\mathbb{C}^{4}\rightarrow\mathbb{C}^{4}$ satisfying
\begin{align}
& \mathcal{J}^{2}=-\bm{1}_{4}, \quad 
\mathcal{J} h(\bm{\mathrm{k}})\mathcal{J}^{-1}=h(\bm{\mathrm{k}})
\end{align}
for every $\bm{\mathrm{k}} \in T^{4}$. 
Then, there must exist a point $\bm{\mathrm{k}}_{*} \in T^{4}$ at which $\det g(\bm{\mathrm{k}}_{*}) = 0$. Equivalently, the quaternionic projector
\begin{align}
& P_{\mathbb{H}}: T^{4} \rightarrow \mathbb{H}P^{1} \cong S^{4}
\end{align}
cannot be an immersion.
\end{theorem}

Theorem \ref{theorem 2} rules out the everywhere nondegeneracy required in Corollary \ref{corollary 1}, but it does not preclude pointwise saturation, as the next result shows.

\begin{theorem} \label{theorem 3}
Consider a gapped four-dimensional four-band Bloch Hamiltonian $h(\bm{\mathrm{k}})$ at half-filling with a fixed antiunitary operator $\mathcal{J}:\mathbb{C}^{4}\rightarrow\mathbb{C}^{4}$ satisfying
\begin{align}
& \mathcal{J}^{2}=-\bm{1}_{4}, \quad 
\mathcal{J} h(\bm{\mathrm{k}})\mathcal{J}^{-1}=h(\bm{\mathrm{k}})
\end{align}
for every $\bm{\mathrm{k}} \in T^{4}$. 
Then, the quaternionic Wirtinger inequality is everywhere saturated 
\begin{align}
& \frac{1}{12}|[\mathrm{Tr}(F\wedge F)]_{1234}(\bm{\mathrm{k}})| = \sqrt{\det g(\bm{\mathrm{k}})}.
\end{align}
\end{theorem}

Combining Theorems \ref{theorem 2} and \ref{theorem 3}, there exists a point $\bm{\mathrm{k}}_{*} \in T^{4}$ at which  
\begin{align}
& \det g(\bm{\mathrm{k}}_{*})=0, \quad [\mathrm{Tr}(F\wedge F)]_{1234}(\bm{\mathrm{k}}_{*})=0.
\end{align}
Thus, pointwise saturation alone does not enable $P_{\mathbb{H}}$ to induce quaternionic Hermitian band geometry on $T^{4}$. Whether the global bound in Corollary \ref{corollary 2} is also saturated is a separate question determined by the sign of the second Chern density.

To illustrate these results explicitly, we consider the minimal four-dimensional lattice Dirac Hamiltonian \cite{QHZ2008LatticeDiracModel}. This model realizes a $\mathcal{J}$-invariant occupied doublet and allows us both to identify the metric-degenerate points predicted by Theorem \ref{theorem 2} and to verify the pointwise saturation at every momentum established in Theorem \ref{theorem 3}. The single-particle Hamiltonian can be written in the compact form $h(\bm{\mathrm{k}}) = d_{a}(\bm{\mathrm{k}}) \Gamma^{a}$ with
\begin{align} \label{lattice dirac model Hamiltonian}
& d(\bm{\mathrm{k}}) = \left[ \sin k_{x}, \sin k_{y}, \sin k_{z}, \sin k_{w}, m+c\sum_{\mu}\cos k_{\mu} \right]
\end{align}
as a five-dimensional vector. Here, the five matrices $\Gamma^{a}$ satisfy the Clifford algebra $\{ \Gamma^{a}, \Gamma^{b} \} = 2\delta^{ab}\bm{1}_{4}$:
\begin{align} \label{gamma representation}
& \Gamma^{i} = -\tau_{2} \otimes \sigma_{i}, \quad
\Gamma^{4} =  \tau_{1} \otimes \bm{1}_{2}, \quad
\Gamma^{5} =  \tau_{3} \otimes \bm{1}_{2}.
\end{align}
The spectrum consists of two doubly degenerate eigenvalues $E_{\pm}(\bm{\mathrm{k}}) = \pm \sqrt{\sum_{a}d_{a}^{2}(\bm{\mathrm{k}})}$. For $c \ne 0$, the bulk gap remains open precisely when 
\begin{align}
& m/c \notin \{ -4,-2,0,2,4 \}.
\end{align}
We assume this condition throughout. The two-fold degeneracy is protected by the $\bm{\mathrm{k}}$-preserving antiunitary symmetry $\mathcal{J} = -(\bm{1}_{2} \otimes i\sigma_{2})\mathcal{K}$ used in the $S^{4}$ coherent-state construction of Sec. \ref{section 2}. As shown in Appendix \ref{appendix E}, for all $\bm{\mathrm{k}}\in T^{4}$ the quaternionic Wirtinger inequality saturates:
\begin{align} \label{saturation of lattice}
& \frac{1}{12}|[\mathrm{Tr}(F \wedge F)]_{1234}| 
= \sqrt{\det g} 
= \frac{1}{4}|\det \mathcal{G}(\bm{\mathrm{k}})|,
\end{align}
where $\mathcal{G} = (\hat{d}, \partial_{x}\hat{d}, \partial_{y}\hat{d}, \partial_{z}\hat{d}, \partial_{w}\hat{d})$ and $\hat{d}_{a} = d_{a}/|d|$. 
At $\bm{\mathrm{k}}_{*} = (\pi/2,\pi/2,0,0)$, the same calculation shows that $\det \mathcal{G}$ vanishes and changes sign across $\bm{\mathrm{k}}_{*}$. Equation (\ref{saturation of lattice}) then gives $\det g(\bm{\mathrm{k}}_{*})=0$, while Eq. (\ref{det g kpm}) shows that the second Chern density likewise changes sign across $\bm{\mathrm{k}}_{*}$. Hence, $P_{\mathbb{H}}$ fails to be immersive, while the sign change makes the global bound of Corollary \ref{corollary 2} strict:
\begin{align}
& \frac{2\pi^{2}}{3}|\mathcal{C}_{2}| < \int_{T^{4}} \sqrt{\det g(\bm{\mathrm{k}})} \mathrm{d}^{4}k.
\end{align}
Thus, the model provides an explicit illustration of Theorems \ref{theorem 2} and \ref{theorem 3} and distinguishes pointwise saturation from both nondegenerate geometric realization and saturation of the global bound. 

We finally ask whether the four-band obstruction can be removed by increasing the number of ambient bands while retaining a single occupied quaternionic band. In two dimensions, the immersion obstruction of a two-band system can be avoided by adding bands and enlarging the target space from $\mathbb{C}P^{1}$ to a higher-dimensional complex projective space. The same strategy, however, does not extend straightforwardly to four dimensions. 
Although additional bands enlarge the target space from $\mathbb{H}P^{1}$ to $\mathbb{H}P^{N-1}$, every projector map satisfying pointwise saturation whose pullback metric is nondegenerate has its image contained in a fixed quaternionic projective line $\mathbb{H}P^{1} \subset \mathbb{H}P^{N-1}$ \cite{Yang2026QKgeometryTRS}. Thus, the effective target remains four-dimensional, and the additional ambient dimensions cannot remove the immersion obstruction. This leads to the following global obstruction.

\begin{theorem} \label{theorem 4}
For $N\ge2$, consider a smooth quaternionic projector map $P_{\mathbb{H}}: T^{4} \rightarrow \mathbb{H}P^{N-1}$ associated with a quaternionic line, and write $g=P_{\mathbb{H}}^{*}g_{\mathrm{FS}}$ for the corresponding quantum metric. Suppose that the quaternionic Wirtinger inequality is saturated at every $\bm{\mathrm{k}} \in T^{4}$. 
Then, there must exist a point $\bm{\mathrm{k}}_{*} \in T^{4}$ at which $\det g(\bm{\mathrm{k}}_{*})=0$. Equivalently, $P_{\mathbb{H}}$ cannot be an immersion. 
\end{theorem}

Theorem \ref{theorem 4} shows that the obstruction is not a peculiarity of the minimal four-band representation. Even when the ambient Hilbert space is enlarged, a single quaternionic band satisfying everywhere saturation cannot use the additional projective directions to evade the obstruction. Thus, the obstruction is insensitive to the number of unoccupied bands as long as the occupied subspace remains a single quaternionic band.

\section{Discussion and Outlook}
\label{section 5}
The central conclusion of this work is that pointwise saturation of the quaternionic Wirtinger inequality does not by itself guarantee a global quaternionic Hermitian band geometry. At metric-nondegenerate points, saturation enforces local quaternionic compatibility, whereas at metric-degenerate points the equality is automatic and carries no quaternionic information. A global realization from pointwise saturation therefore additionally requires an everywhere nondegenerate pullback metric, or equivalently, an immersive projector map. 
This distinction follows directly from the projector-based relation between the non-Abelian Berry curvature and quantum metric of a single quaternionic band. For the minimal-charge LLL on $S^{4}$, the metric is everywhere nondegenerate and the inequality is saturated, so the projector induces quaternionic Hermitian band geometry.
On $T^{4}$, by contrast, everywhere saturation necessarily coexists with metric-degenerate points, and this obstruction persists after adding additional unoccupied bands as long as the occupied sector remains a single quaternionic band. Nevertheless, wherever the metric is nondegenerate, the saturated $T^{4}$ geometry still exhibits the same local quaternionic compatibility as on $S^{4}$.

Despite this distinction between local compatibility and global realization, the local geometry is still experimentally meaningful. Experimental access to the Abelian quantum geometry~\cite{Pump2D1,Pump2D2} and recent advances in its non-Abelian counterpart~\cite{NonAbelianQGT1,NonAbelianQGT2} motivate extending such measurements to the non-Abelian setting. 
Beyond such local diagnostics, the existence of everywhere nondegenerate saturation provides a finer geometric criterion within a given second Chern sector. For a closed connected parameter space, this condition further locks geometry to topology through
\begin{align}
& \int_{X} \sqrt{\det g}\ \mathrm{d}^{4}x = \frac{2\pi^{2}}{3}|\mathcal{C}_{2}|.
\end{align}
Such global locking provides a complementary geometric characterization of the second Chern topology: the magnitude of the second Chern number is fixed by the total quantum volume, while the quantum geometry can still distinguish different band realizations within the same topological sector.

To illustrate how the global obstruction may be avoided, let $p \in S^{4}$ and consider the stereographic parameterization
\begin{align}
& \sigma: \mathbb{R}^{4} \rightarrow S^{4}\backslash\{p\}.
\end{align}
The pullback projector $P_{\mathbb{H}} \circ \sigma: \mathbb{R}^{4}\rightarrow \mathbb{H}P^{1}$ is an immersion and hence induces a quaternionic Hermitian band geometry on $\mathbb{R}^{4}$. The stereographic pullback of the associated Yang-monopole Berry connection on $S^{4}$ is gauge-equivalent to the BPST instanton connection on $\mathbb{R}^{4}$ \cite{Hasebe2020RelationbtwRandS,Hasebe2020YangMonopoleAndBPSTInstanton}. The induced quantum metric is the stereographic pullback of the rescaled round metric on $S^{4}$, and is therefore conformally flat but not equal to the flat Euclidean metric on $\mathbb{R}^{4}$. 

The sharp contrast between $T^{4}$ and $\mathbb{R}^{4} \cong S^{4}\backslash\{p\}$ naturally suggests the interpolating family
\begin{align}
& X_{a} = T^{4-a} \times \mathbb{R}^{a}, \quad a=0,1,2,3,4.
\end{align}
For $a \ge 1$, decompactification removes the compactness argument underlying the $T^{4}$ obstruction, although it does not by itself guarantee a physically admissible realization. It remains to determine whether a gapped mixed lattice-continuum Hamiltonian can induce quaternionic Hermitian band geometry globally on $X_{a}$. The minimally decompactified space $T^{3}\times\mathbb{R}$ provides a natural starting point, where self-dual Yang-Mills configurations have been studied under suitable asymptotic or twisted boundary conditions \cite{VanBaal1996InstantonT3R1, VanBaal1984SUNYangMillswithConstantFieldonT4,VanBaal1999NahmGaugeFieldforTorus}.
Synthetic dimensions may provide a route to realizing such mixed parameter spaces and probing four-dimensional second Chern physics \cite{Pump4D1,Pump4D2,Pump4D3}. Establishing such models would determine whether decompactification allows pointwise saturation and metric nondegeneracy to coexist everywhere on $X_{a}$, thereby realizing a global quaternionic Hermitian band geometry.
\\

\textit{Note added.}---While this manuscript was nearing completion, two related works appeared independently~\cite{Yang2026QKgeometryTRS,Zhao2026SecondChernBoundsNonAbelian}, both deriving an equivalent inequality. Specifically, Ref.~\cite{Zhao2026SecondChernBoundsNonAbelian} obtained the inequality via Hodge decomposition and independently established the four-band obstruction on $T^4$. Ref.~\cite{Yang2026QKgeometryTRS} formulated the inequality within quaternionic projective geometry and established the local $\mathbb{H}P^1$ rigidity using their Proposition~S7. All results in the present work were derived independently, with the sole exception of Theorem~\ref{theorem 4}, which utilizes Proposition~S7 of Ref.~\cite{Yang2026QKgeometryTRS}. 
See Appendix \ref{appendix F} for differences in the usage of the term ``quaternionic K\"ahler geometry'' and for a comparison of the explicit geometric conditions in Refs.~\cite{Yang2026QKgeometryTRS,Zhao2026SecondChernBoundsNonAbelian} with the definitions adopted in the present work.

%{\color{red} See Appendix F for the different usage of the definitions for ``quaternionic Kahler geometry'', algebraic definition in ... and diff geom. definition used in this paper. (GYC: add appropriate details at Appendix F and revise this sentence accordingly.)}  

\begin{acknowledgements}
We thank Hee-Cheol Kim, Bruno Mera, and Tomoki Ozawa for helpful discussions. D.~H., M.~Y., and G.~Y.~C. acknowledge the financial supports by Samsung Science and Technology Foundation under Project Number SSTF-BA2401-03, the NRF of Korea (Grants No. RS-2026-25479545, RS-2024-00410027, RS-2023-NR119931, RS-2024-00444725, RS-2023-00256050, IRS-2025-25453111, RS-2025-08542968) funded by the Korean Government (MSIT), the Air Force Office of Scientific Research under Award No. FA23862514026, and Institute of Basic Science under project code IBS-R014-D1. D.~H, and M.~Y.. acknowledge the support from Agency for Defense Development through a grant funded by the Defense Acquisition Program Administration (DAPA) (UI257011TE). 
\end{acknowledgements}

\appendix

\section{Quaternionic Linear Algebra and Geometry}
\label{appendix A}
\subsection{Quaternionic Vector Spaces}
A quaternion is written as \cite{Rodman2014TopicsInQuaternionLinearAlgebra, Adler1996QuaternionicQuantumMechanics}
\begin{align}
& q = q_{0} + q_{1}e_{1} + q_{2}e_{2} + q_{3}e_{3}, \quad q_{i} \in \mathbb{R},
\end{align}
where the imaginary quaternionic units satisfy $e_{i}e_{j}=-\delta_{ij}+\epsilon_{ijk}e_{k}$ for $i,j=1,2,3$. Quaternionic conjugation and the norm are defined by 
\begin{align}
& \bar{q} = q_{0}-q_{a}e_{a}, \quad |q|^{2} = \bar{q}q = q_{0}^{2} + q_{1}^{2} + q_{2}^{2} + q_{3}^{2}.
\end{align}
The quaternionic algebra admits an equivalent complex $2\times2$ matrix representation. Under the identification $e_{i} \leftrightarrow -i\sigma_{i}$, where $\sigma_{i}$ are the Pauli matrices, a quaternion is represented by
\begin{align}
& \rho(q) = q_{0}\bm{1}_{2} - iq_{i}\sigma_{i}.
\end{align}
Consequently, the unit quaternions form the group
\begin{align}
& \mathrm{Sp}(1) = \{ q\in\mathbb{H}; |q|=1 \} \cong \mathrm{SU}(2).
\end{align}

A quaternionic vector space may be defined using either left or right scalar multiplication. Throughout this work, we regard $V= \mathbb{H}^{N}$ as a right $\mathbb{H}$-module and adopt the convention
\begin{align}
& V \times \mathbb{H} \rightarrow V, \quad 
(\Psi,\lambda)\mapsto \Psi\lambda.
\end{align}
Accordingly, all quaternionic-linear maps are understood to be right $\mathbb{H}$-linear. In particular, if $A\in \mathrm{End}(V)$, a nonzero vector $\Psi\in V$ is a right eigenvector with eigenvalue $q \in \mathbb{H}$ when $A\Psi = \Psi q$.
With this convention, the standard quaternionic Hermitian inner product is 
\begin{align}
& \langle \Psi, \Phi\rangle_{\mathbb{H}}
= \sum_{i=1}^{N} \overline{\Psi}_{i}\Phi_{i},
\end{align}
and satisfies $\langle \Psi, \Phi\rangle_{\mathbb{H}} = \overline{\langle \Phi, \Psi\rangle_{\mathbb{H}}}$. The quaternionic unitary group is then defined by 
\begin{align}
& \mathrm{Sp}(N) = \{ U\in \mathrm{Mat}_{N\times N}(\mathbb{H}) ; U^{\dagger}U = \bm{1}_{N} \},
\end{align}
which acts transitively on the unit sphere $S^{4N-1} \subset \mathbb{H}^{N}$ while preserving the quaternionic Hermitian product. 

Analogously to its real and complex counterparts, the quaternionic projective space $\mathbb{H}P^{N-1}$ is defined as the set of quaternionic lines in $\mathbb{H}^{N}$. We have the canonical projection
\begin{align}
& \pi:\mathbb{H}^{N}\backslash\{0\} \rightarrow \mathbb{H}P^{N-1}:x\mapsto\pi(x)=[x]=x\mathbb{H}.
\end{align}
Equivalently, quaternionic projective space can be represented as the homogeneous space \cite{besse2007EinsteinManifolds}
\begin{align}
& \mathbb{H}P^{N-1} \cong \frac{\mathrm{Sp}(N)}{\mathrm{Sp}(N-1)\times\mathrm{Sp}(1)}.
\end{align}
A normalized local representative may be chosen as
\begin{align}
& \Psi(w) = \frac{1}{\sqrt{1+|w|^{2}}}\begin{pmatrix} 1 \\  w \end{pmatrix}.
\end{align}
In the inhomogeneous coordinates $w\in \mathbb{H}^{N-1}$, the metric takes the form 
\begin{align}
& \mathrm{d}s^{2}_{\mathrm{FS}} = \frac{2(1+w^{\dagger}w)\mathrm{d}w^{\dagger}\mathrm{d}w - 2(\mathrm{d}w^{\dagger}w)(w^{\dagger}\mathrm{d}w)}{(1+w^{\dagger}w)^{2}}.
\end{align}
In particular, for $\mathbb{H}P^{1}$,
\begin{align}
& \mathrm{d}s^{2}_{\mathrm{FS}} = \frac{2\mathrm{d}\bar{w}\mathrm{d}w}{(1+|w|^{2})^{2}},
\end{align}
which is one half of the standard round metric on $S^{4}$ expressed in stereographic coordinates.

\subsection{Definitions and Conventions in Quaternionic Geometry}
For completeness, we collect below the definitions and dimension-dependent conventions for quaternionic geometry adopted throughout this work 
\cite{Swann1990QK,Alekseevsky1999AlmostisQuaternonic,Alekseevsky1996quaternionic,besse2007EinsteinManifolds,Salamon198QM,Salamon1982QKM}.

\begin{definition} \label{definition 1}
An \textbf{almost quaternionic structure} on a smooth manifold $X$ is a rank-three vector subbundle $\mathcal{Q} \subset \mathrm{End}(TX)$, which is locally spanned by an admissible frame $\mathcal{Q}|_{U} = \mathrm{span}\{I^{1},I^{2},I^{3}\}$ satisfying 
\begin{align}
& I^{a}I^{b} = -\delta^{ab}\mathrm{id}_{TX} + \epsilon^{abc}I^{c}, \quad a,b=1,2,3.
\end{align}
For each $x \in X$, the fiber $\mathcal{Q}_{x}\subset \mathrm{End}(T_{x}X)$ determines a linear quaternionic structure on $T_{x}X$.
\end{definition}

\begin{definition}
Let $X$ be a smooth manifold of real dimension $4n$, with $n\ge1$. If $X$ is equipped with an almost quaternionic structure $\mathcal{Q}$, then the pair $(X,\mathcal{Q})$ is called an \textbf{almost quaternionic manifold}.
\end{definition}

The individual endomorphisms $I^{i}$ need not be globally defined. Two admissible frames on overlapping coordinate patches are related by 
\begin{align} \label{definition 2}
& \tilde{I}^{i} = R_{ij}I^{j}, \quad R:U \cap \widetilde{U} \rightarrow \mathrm{SO}(3).
\end{align}
Consequently, the rank-three bundle $\mathcal{Q}$ is globally defined even when no individual $I^{i}$ is global.

\begin{definition} \label{definition 3}
An almost quaternionic structure $\mathcal{Q}$ on a smooth manifold $X$ is called a \textbf{quaternionic structure} if there exists a torsion-free affine connection $\nabla$ on $TX$ preserving $\mathcal{Q}$:
\begin{align}
& \nabla \mathcal{Q} \subset \mathcal{Q}.
\end{align}
\end{definition}

In real dimension four, an almost quaternionic structure is equivalent to an oriented conformal structure and always admits a torsion-free connection preserving $\mathcal{Q}$ \cite{Alekseevsky1999AlmostisQuaternonic,Ivanov2002AlmostIsQuaternioinic}. Hence, every four-dimensional almost quaternionic structure is automatically quaternionic in the sense of Definition \ref{definition 3}. By contrast, for $n\ge2$, the existence of such a connection is a nontrivial differential condition. This provides a low-dimensional analogue of the situation in complex geometry, where every almost complex structure on a real two-dimensional manifold is automatically integrable and hence defines a complex structure. Since all quaternionic parameter manifolds considered in this work are four-dimensional, we suppress the qualifier ``almost'' throughout and refer to the induced structures simply as quaternionic structures. 

\begin{definition} \label{definition 4}
Let $(X,\mathcal{Q})$ be an almost quaternionic manifold of real dimension $4n$, with $n\ge1$, and let $g$ be a Riemannian metric on $X$. The metric $g$ is called compatible with $\mathcal{Q}$ if every local admissible frame $\{I^{1},I^{2},I^{3}\}$ acts orthogonally with respect to $g$:
\begin{align}
& g(I^{i}u,I^{i}v) = g(u,v), \quad i=1,2,3
\end{align}
for all $x\in X$, $u,v\in T_{x}X$. If $g$ is compatible with $\mathcal{Q}$, then the triple $(X,g,\mathcal{Q})$ is called a \textbf{quaternionic Hermitian manifold}.
\end{definition}

This is a pointwise compatibility condition and does not require the existence of a torsion-free connection preserving $\mathcal{Q}$. The corresponding local fundamental 2-forms are 
\begin{align} \label{relation btw w and g}
& \omega^{i}(u,v) = g(I^{i}u,v), \quad i=1,2,3.
\end{align}
Under a change of admissible frame $\tilde{I}^{i}=R^{ij}I^{j}$, with $R \in \mathrm{SO}(3)$, the 2-forms transform in the same way $\widetilde{\omega}^{i}=R^{ij}\omega^{j}$. Hence, although the individual $\omega^{i}$ are generally defined only locally, their rank-three span is globally well-defined. 

\begin{definition} \label{definition 5}
Let $(X,g,\mathcal{Q})$ be a quaternionic Hermitian manifold. The \textbf{fundamental 4-form} associated with $(g,\mathcal{Q})$ is defined by
\begin{align}
& \Omega = \frac{1}{6}(\omega^{1} \wedge \omega^{1} + \omega^{2} \wedge \omega^{2} + \omega^{3} \wedge \omega^{3}).
\end{align}
\end{definition}

The $\mathrm{SO}(3)$ transformation law above implies that this expression is independent of the choice of local admissible frame:
\begin{align} 
& \sum_{i} \widetilde{\omega}^{i} \wedge \widetilde{\omega}^{i}
= \sum_{i,j,k} R^{ij}R^{ik} \omega^{j} \wedge \omega^{k}
= \sum_{i} \omega^{i} \wedge \omega^{i}.
\end{align}
Therefore, $\Omega$ is a globally defined differential 4-form, known as the Kraines form \cite{Kraines1966TopologyOfQuaternionicManifolds}. It is the quaternionic counterpart of the K\"ahler form and provides the fundamental differential form entering the quaternionic Wirtinger inequality.

\begin{definition} \label{definition 6}
Let $(X,g,\mathcal{Q})$ be a quaternionic Hermitian manifold of real dimension $4n$, with $n \ge 2$. It is called a \textbf{quaternionic K\"ahler manifold} if its Levi-Civita connection $\nabla^{g}$ preserves $\mathcal{Q}$:
\begin{align}
& \nabla^{g}\mathcal{Q} \subset \mathcal{Q}.
\end{align}
Equivalently, its Riemannian holonomy satisfies 
\begin{align}
& \mathrm{Hol}(g) \subset \mathrm{Sp}(n)\cdot\mathrm{Sp}(1).
\end{align}
In terms of the fundamental 4-form $\Omega$, these conditions are also equivalent to 
\begin{align}
& \nabla^{g}\Omega = 0.
\end{align}
\end{definition}

For $n=1$, the additional conditions entering Definition \ref{definition 6} become automatic once a quaternionic Hermitian structure is given. 
With the normalization adopted here, $\Omega$ coincides with the Riemannian volume form, so $\nabla^{g}\Omega=0$; equivalently, 
\begin{align}
& \mathrm{Hol}(g) \subset \mathrm{Sp}(1)\cdot\mathrm{Sp}(1)\cong\mathrm{SO}(4)
\end{align}
imposes no further restriction. 
Thus, a literal extension of Definition \ref{definition 6} to $n=1$ would identify every four-dimensional quaternionic Hermitian manifold as quaternionic K\"ahler. 
To retain a nontrivial four-dimensional notion, we therefore adopt the following curvature-based convention.

%\begin{definition} \label{definition 7}
%A four-dimensional quaternionic Hermitian manifold $(X,g,\mathcal{Q})$ is called a \textbf{quaternionic K\"ahler manifold} if $g$ is Einstein and self-dual:
%\begin{align}
%& \mathrm{Ric}_{g}=\kappa g, \quad W^{-}=0,
%\end{align}
%for some constant $\kappa$, where $W^{-}$ denotes the anti-self-dual Weyl curvature with respect to the orientation determined by $\mathcal{Q}$.
%\end{definition}

\begin{definition} \label{definition 7}
An oriented Riemannian four-manifold $(X,g)$ is called a \textbf{quaternionic K\"ahler manifold} if $g$ is Einstein and self-dual:
\begin{align}
& \mathrm{Ric}_{g}=\kappa g, \quad W^{-}=0,
\end{align}
for some constant $\kappa$, where $W^{-}$ denotes the anti-self-dual Weyl curvature with respect to the chosen orientation.
\end{definition}

In real dimension four, the compatibility condition defining quaternionic Hermitian geometry remains nontrivial for a fixed quaternionic structure $\mathcal{Q}$, since a Riemannian metric $g$ is still required to be compatible with $\mathcal{Q}$.
For a four-dimensional quaternionic Hermitian manifold $(X,g,\mathcal{Q})$, the structure $\mathcal{Q}$ determines an orientation on $X$, and the manifold is quaternionic K\"ahler in the sense of Definition \ref{definition 7} if and only if $g$ is additionally Einstein and self-dual with respect to this orientation. These additional curvature conditions provide a nontrivial four-dimensional counterpart of higher-dimensional quaternionic K\"ahler geometry \cite{Swann1990QK}.

\section{Auxiliary Results and Proofs}
\label{appendix B}

\subsection{Auxiliary Lemmas}
\begin{lemma} \label{lemma 1}
Let $V_{\mathbb{C}} \cong \mathbb{C}^{2N}$ be a complex Hermitian vector space equipped with an antiunitary operator satisfying $\mathcal{J}^{2}=-\bm{1}_{2N}$. The underlying real vector space of $V_{\mathbb{C}}$ then carries a natural quaternionic vector space structure $V_{\mathbb{H}} \cong \mathbb{H}^{N}$, where multiplication by $i$ is given by the original complex structure and the action of $j$ is given by $\mathcal{J}$.
Then, there is a natural diffeomorphism of real smooth manifolds
\begin{align*}
& \eta_{r,N}: \mathrm{Gr}_{r,N}^{\mathbb{H}} \overset{\cong}{\longrightarrow} (\mathrm{Gr}_{2r,2N}^{\mathbb{C}})^{\mathcal{J}} 
\end{align*}
obtained by regarding a quaternionic $r$-plane as its underlying complex $2r$-plane. Moreover, if $P_{V}$ denotes the orthogonal projector onto a complex subspace $V \subset V_{\mathbb{C}}$, then 
\begin{align*}
& \mathcal{J}V=V \Longleftrightarrow \mathcal{J}P_{V}\mathcal{J}^{-1}=P_{V}.
\end{align*}
\end{lemma}

$\quad Proof.$ 
Under the right-multiplication convention, let $\mathcal{I}$ denote the original complex structure on $V_{\mathbb{C}}$, acting by $\mathcal{I}v = vi$ for $v \in V_{\mathbb{C}}$. The antiunitary operator $\mathcal{J}$ acts by $\mathcal{J}v = vj$. For every $\lambda \in \mathbb{C}_{i}:=\mathrm{span}\{1,i\}\subset \mathbb{H}$, the quaternionic identity $\lambda j = j \bar{\lambda}$ gives 
\begin{align}
& \mathcal{J}(v\lambda) = (v\lambda)j = v(\lambda j) = v(j\bar{\lambda}) = (\mathcal{J}v)\bar{\lambda}.
\end{align}
Consequently, we obtain $\mathcal{IJ}=-\mathcal{JI}$:
\begin{align}
& (\mathcal{IJ})v = \mathcal{I}(vj) 
= v(ji)  = -v(ij) = -(\mathcal{JI})v.
\end{align}
Indeed, $v(ij) = (\mathcal{JI})v = vk$. Thus, the underlying real vector space of $V_{\mathbb{C}}$ acquires a natural right quaternionic vector space structure $V_{\mathbb{H}} \cong \mathbb{H}^{N}$. 

Let $L \in \mathrm{Gr}_{r,N}^{\mathbb{H}}$. Regarding $L$ as a complex vector space gives a complex subspace of $V_{\mathbb{C}}$ of dimension $2r$. Since $L$ is closed under right multiplication by $j$, it is invariant under $\mathcal{J}$. Therefore,
\begin{align}
& \eta_{r,N}(L) \in (\mathrm{Gr}_{2r,2N}^{\mathbb{C}})^{\mathcal{J}}.
\end{align}
Conversely, let $W \in (\mathrm{Gr}_{2r,2N}^{\mathbb{C}})^{\mathcal{J}}$. Because $W$ is a complex subspace, it is closed under multiplication by $i$, and because $\mathcal{J}W=W$, it is also closed under multiplication by $j$. It is consequently closed under multiplication by $k=ij$, and hence by every quaternion. Thus, $W$ is a quaternionic subspace of $V_{\mathbb{H}}$. Then, $\eta_{r,N}$ is a bijection. 
The correspondence leaves the underlying vectors unchanged and is induced by the fixed real-linear identification $\mathbb{H}^{N}\cong\mathbb{C}^{2N}$. The fixed linear identification induces smooth maps between the corresponding Grassmannians, with smooth inverse. Hence, $\eta_{r,N}$ is a diffeomorphism of real smooth manifolds.

Finally, suppose that $\mathcal{J}V=V$. Since $\mathcal{J}$ is antiunitary, it also preserves $V^{\perp}$, and hence preserves the orthogonal decomposition
\begin{align}
& V_{\mathbb{C}} = V \oplus V^{\perp}.
\end{align}
For $x=v+w$, where $v\in V$ and $w\in V^{\perp}$, we then have
\begin{align}
& P_{V}(\mathcal{J}x) = P_{V}(\mathcal{J}v + \mathcal{J}w)
= \mathcal{J}v = \mathcal{J}(P_{V}x).
\end{align}
Conversely, suppose that $\mathcal{J}P_{V}\mathcal{J}^{-1}=P_{V}$. For every $v\in V$, 
\begin{align}
& P_{V}(\mathcal{J}v) = \mathcal{J}(P_{V}v) = \mathcal{J}v.
\end{align}
Thus, $\mathcal{J}v\in V$, so $\mathcal{J}V\subset V$. Since $\mathcal{J}$ is invertible, this inclusion is an equality. $\blacksquare$ \\

In the band-theoretic setting, $V_{\mathbb{C}}$ represents the ambient complex single-particle Hilbert space, while $V \subset V_{\mathbb{C}}$ is the occupied subspace at a fixed point of the parameter space. Its orthogonal complement $V^{\perp}$ represents the unoccupied sector, and $P_{V}$ is the corresponding spectral projector onto $V$. 
Lemma \ref{lemma 1} translates the invariance condition $\mathcal{J}V=V$ into the gauge-invariant projector condition $\mathcal{J}P_{V}\mathcal{J}^{-1}=P_{V}$, which states that $\mathcal{J}$ does not mix the occupied and unoccupied sectors. 
Thus, the orthogonal projector defines a quaternionic band. Complex and quaternionic bands can therefore be treated within the same projector-based framework, allowing the usual projector formulation of complex band geometry to be extended to the quaternionic setting.
This characterization also provides the starting point for the factorization of $P_{V}$ through a quaternionic projector $P_{\mathbb{H}}$, as established in Lemma \ref{lemma 2}.

\begin{lemma} \label{lemma 2}
Let $P: X \rightarrow \mathrm{Gr}_{2,2N}^{\mathbb{C}}$ be a smooth complex rank-two projector, and let $E$ be its image bundle, with fibers $E_{x} = \mathrm{Im}P(x)$. 
Suppose that there exists an antiunitary operator $\mathcal{J}$ on $\mathbb{C}^{2N}$ such that
\begin{align}
& \mathcal{J}^{2}= -\bm{1}_{2N}, \quad \mathcal{J}P(x)\mathcal{J}^{-1}=P(x)
\end{align}
for every $x \in X$. Then, $E$ is naturally a quaternionic line bundle, and $P$ factors uniquely through the quaternionic Grassmannian as 
\begin{align}
& P_{\mathbb{H}}: X \rightarrow \mathrm{Gr}_{1,N}^{\mathbb{H}} \cong \mathbb{H}P^{N-1},
\end{align}
satisfying $P = \eta_{1,N} \circ P_{\mathbb{H}}$.
\end{lemma}

$\quad Proof.$ 
Since $P$ is smooth and has complex rank-two, its image $E = \mathrm{Im}P$ is a smooth complex rank-two vector bundle over $X$.
For any $v \in E_{x}$, we have $P(x)v =v$, and hence 
\begin{align}
& P(x)(\mathcal{J}v) = \mathcal{J}(P(x)v) = \mathcal{J}v.
\end{align}
Therefore, $\mathcal{J}v \in E_{x}$, so that $\mathcal{J}E_{x} \subset E_{x}$. Since $\mathcal{J}$ is invertible, this inclusion is an equality. 
Moreover, $E_{x}$ is closed under right multiplication by $i$ because it is a complex subspace, and it is closed under right multiplication by $j$ because $\mathcal{J}v=vj$. It is therefore closed under right multiplication by $k=ij$, and hence by every quaternion. Since $\dim_{\mathbb{C}}E_{x}=2$, it follows that $\dim_{\mathbb{H}} E_{x} = 1$, and hence each $E_{x}$ is a quaternionic line. 
Because the complex bundle $E$ is smooth and the operator $\mathcal{J}$ is fixed on $V_{\mathbb{C}}$, this quaternionic scalar multiplication varies smoothly over $X$. Hence, $E$ is a quaternionic line bundle. Equivalently, $P(x) \in (\mathrm{Gr}_{2,2N}^{\mathbb{C}})^{\mathcal{J}}$.

By Lemma \ref{lemma 1}, the diffeomorphism $\eta_{1,N}$ allows us to define
\begin{align}
& P_{\mathbb{H}} := \eta_{1,N}^{-1} \circ P: X \rightarrow \mathrm{Gr}_{1,N}^{\mathbb{H}} \cong \mathbb{H}P^{N-1}.
\end{align}
Since $P$ and $\eta_{1,N}^{-1}$ are smooth, $P_{\mathbb{H}}$ is smooth, and we obtain $P=\eta_{1,N}\circ P_{\mathbb{H}}$. Finally, the factorization is unique because $\eta_{1,N}$ is injective. $\blacksquare$ \\

Physically, Lemma \ref{lemma 2} upgrades the pointwise $\mathcal{J}$-pairing of the occupied states to a global quaternionic band structure. The factorization $P_{\mathbb{H}}=\eta^{-1}_{1,N}\circ P$ shows that the projector $P$ takes values not in the full complex Grassmannian but in its quaternionic locus $\mathbb{H}P^{N-1}$. Consequently, the structure group of the occupied doublet reduces from $\mathrm{U}(2)$ to $\mathrm{SU}(2)\cong\mathrm{Sp}(1)$, so that the associated Berry connection and curvature are naturally non-Abelian $\mathrm{SU}(2)$ gauge fields. This global reduction provides the band-theoretic foundation for the quaternionic quantum geometry considered below.

\begin{lemma} \label{lemma 3}
Let $\mathcal{S} \rightarrow \mathbb{H}P^{N-1}$ be the tautological quaternionic line bundle equipped with the universal $\mathrm{SU}(2)\cong\mathrm{Sp}(1)$-Berry connection induced from the Hopf fibration $S^{4N-1} \rightarrow \mathbb{H}P^{N-1}$, and let $\mathcal{F}$ denote its Berry curvature. 
Let $\mathcal{Q}_{\mathrm{FS}} \subset \mathrm{End}(T\mathbb{H}P^{N-1})$ be the canonical quaternionic structure. On an open set $U \subset \mathbb{H}P^{N-1}$, choose a local admissible frame 
\begin{align}
& \mathcal{Q}_{\mathrm{FS}}|_{U} = \mathrm{span}\{ I^{1}_{\mathrm{FS}}, I^{2}_{\mathrm{FS}}, I^{3}_{\mathrm{FS}} \},
\end{align}
and use the corresponding local $\mathrm{SU}(2)$ gauge frame represented by $T_{i}=-i\sigma_{i}/2$ for $i=1,2,3$. 
Define the local fundamental 2-forms by
\begin{align}
& \omega_{\mathrm{FS}}^{i}(v,w) = g_{\mathrm{FS}}(I^{i}_{\mathrm{FS}}v,w), \
v,w\in T_{u}\mathbb{H}P^{N-1}
\end{align}
for $u\in U$. Writing the Berry curvature locally as $\mathcal{F}=\mathcal{F}^{i}T_{i}$, we obtain
\begin{align}
& \mathcal{F}^{i} = -2\omega^{i}_{\mathrm{FS}}, \quad i=1,2,3.
\end{align}
\end{lemma}

$\quad Proof.$ 
Let $\pi$ denote the quaternionic Hopf projection
\begin{align}
& \pi: S^{4N-1} \rightarrow \mathbb{H}P^{N-1}, \quad \pi(x) = [x].
\end{align}
Fix $u:=[x] \in U$, with normalized representative $x\in S^{4N-1}$, and let $v,w\in T_{[x]}\mathbb{H}P^{N-1}$. Let $\widetilde{v},\widetilde{w}$ denote their horizontal lifts at $x$, characterized by 
\begin{align}
& \mathrm{d}\pi_{x}(\widetilde{v}) = v, \quad
\mathrm{d}\pi_{x}(\widetilde{w}) = w.
\end{align}
Write their quaternionic Hermitian inner product as
\begin{align}
& \langle \widetilde{v},\widetilde{w}\rangle_{\mathbb{H}}
= q_{0} + \sum_{i=1}^{3}q_{i}e_{i}, \quad q_{i} \in \mathbb{R}.
\end{align}
In the complex rank-two normalization, the quaternionic Fubini-Study metric is 
\begin{align}
& g_{\mathrm{FS}}(v,w) = 2\mathrm{Re}[\langle \widetilde{v},\widetilde{w}\rangle_{\mathbb{H}}] = 2q_{0}.
\end{align}

Under the right-multiplication convention, we define endomorphisms of $T_{[x]}\mathbb{H}P^{N-1}$ by
\begin{align}
& I^{i}_{\mathrm{FS}}v := \mathrm{d}\pi_{x}( -\widetilde{v}e_{i}).
\end{align}
This is well-defined because right multiplication preserves the horizontal subspace:
\begin{align}
& \langle x,\widetilde{v}e_{i}\rangle_{\mathbb{H}}
= \langle x,\widetilde{v}\rangle_{\mathbb{H}}e_{i} = 0.
\end{align}
Moreover, for the chosen local Hopf section, $e_{i}e_{j} = -\delta_{ij} + \epsilon_{ijk}e_{k}$ directly verifies that 
\begin{align}
& I^{i}_{\mathrm{FS}}I^{j}_{\mathrm{FS}} = -\delta^{ij} \mathrm{id}_{TU} + \epsilon^{ijk}I^{k}_{\mathrm{FS}}.
\end{align}
Thus, $\{I^{i}_{\mathrm{FS}}\}_{i=1}^{3}$ is a local admissible quaternionic frame. A change of representative $x \mapsto xq$ for $ q\in\mathrm{Sp}(1)$ leaves $\mathcal{Q}_{\mathrm{FS}}$ unchanged. 
Using the quaternionic Hermitian identity $\langle v\lambda,w \rangle_{\mathbb{H}}
= \bar{\lambda} \langle v,w\rangle_{\mathbb{H}}$, together with 
\begin{align}
& \widetilde{I^{i}_{\mathrm{FS}}v} = - \widetilde{v}e_{i},
\end{align}
we obtain $\omega_{\mathrm{FS}}^{i}(v,w)$ by
\begin{align}
g_{\mathrm{FS}}(I^{i}_{\mathrm{FS}}v,w)
& = 2\mathrm{Re} [\langle -\widetilde{v}e_{i},\widetilde{w}\rangle_{\mathbb{H}}]
= 2\mathrm{Re} [e_{i} \langle \widetilde{v},\widetilde{w}\rangle_{\mathbb{H}}] \nonumber \\
& = 2\mathrm{Re}[ e_{i}(q_{0} + \Sigma_{j}q_{j}e_{j}) ] = -2q_{i}.
\end{align}

Next, consider the canonical quaternionic connection one-form of the Hopf fibration and its curvature
\begin{align}
& \mathcal{A}_{\mathbb{H}} = \langle x, \mathrm{d}x\rangle_{\mathbb{H}}, \quad
\mathcal{F}_{\mathbb{H}} = \mathrm{d}\mathcal{A}_{\mathbb{H}} + \mathcal{A}_{\mathbb{H}} \wedge \mathcal{A}_{\mathbb{H}}.
\end{align}
Because $\widetilde{v}$ and $\widetilde{w}$ are horizontal,
\begin{align}
& \mathcal{A}_{\mathbb{H}}(\widetilde{v}) = \langle x,\widetilde{v}\rangle_{\mathbb{H}} = 0, \quad 
\mathcal{A}_{\mathbb{H}}(\widetilde{w})=0.
\end{align}
Therefore, the quadratic term vanishes when evaluated on these horizontal lifts, and 
\begin{align}
\mathcal{F}_{\mathbb{H}}(v,w)
& = \mathrm{d}\mathcal{A}_{\mathbb{H}}(\widetilde{v},\widetilde{w}) 
= \sum_{\alpha=1}^{N}(\mathrm{d}\bar{x}_{\alpha} \wedge \mathrm{d}x_{\alpha}) (\widetilde{v},\widetilde{w}) \nonumber \\
& = \sum_{\alpha=1}^{N} \Bigr(\mathrm{d}\bar{x}_{\alpha}(\widetilde{v}) \mathrm{d}x_{\alpha}(\widetilde{w}) - \mathrm{d}\bar{x}_{\alpha}(\widetilde{w}) \mathrm{d}x_{\alpha}(\widetilde{v}) \Bigr) \nonumber \\ 
& = \langle \widetilde{v},\widetilde{w}\rangle_{\mathbb{H}} - \langle \widetilde{w},\widetilde{v}\rangle_{\mathbb{H}}
= 2 \sum_{i=1}^{3} q_{i}e_{i},
\end{align}
where $x = (x_{1},\ldots,x_{N})^{T} \in \mathbb{H}^{N}$. Under the complex representation $\rho(e_{i}) = -i\sigma_{i} = 2T_{i}$, the corresponding complex Berry curvature is 
\begin{align}
& \mathcal{F}(v,w) = \rho \bigr( \mathcal{F}_{\mathbb{H}}(\widetilde{v},\widetilde{w})\bigr)
= 4 \sum_{i=1}^{3} q_{i}T_{i}.
\end{align}
Since $F^{i}(v,w) = 4q_{i}$ with $\omega_{\mathrm{FS}}^{i}(v,w)=-2q_{i}$, we obtain
\begin{align}
& \mathcal{F}^{i} = -2\omega_{\mathrm{FS}}^{i}, \quad i=1,2,3. 
\quad \blacksquare
\end{align}
\\

Lemma \ref{lemma 3} is the quaternionic counterpart of the familiar relation between the Abelian Berry curvature and the K\"ahler form in complex-band geometry \cite{Mera2021KahlerBandLL}: 
\begin{align}
& \mathcal{F} = -2\omega_{\mathrm{FS}}.
\end{align}

\begin{lemma} \label{lemma 4}
Let $X$ be an $n$-dimensional smooth manifold, let $(Y,h)$ be a Riemannian manifold, and let $f:X\rightarrow Y$ be a smooth map. Let $g=f^{*}h$ be the pullback symmetric 2-tensor. Then, for every $x \in X$, the following conditions are equivalent:
\begin{flalign}
& \text{$\mathrm{(i)}$ $\mathrm{d}f_{x}$ is injective,} \quad \quad \quad \ \ 
\text{$\mathrm{(ii)}$ $\mathrm{rank}\ \mathrm{d}f_{x}=n$,} \nonumber \\
& \text{$\mathrm{(iii)}$ $g_{x}$ is nondegenerate,} \quad
 \text{$\mathrm{(iv)}$ $\det g(x) >0$,} \nonumber &&
\end{flalign}
in any local coordinate system. In particular, $f$ is an immersion if and only if $g$ is nondegenerate everywhere on $X$.
\end{lemma}

$\quad Proof.$ 
Since $g=f^{*}h$, for every $v \in T_{x}X$,
\begin{align}
& g_{x}(v,v) = h_{f(x)}(\mathrm{d}f_{x}(v),\mathrm{d}f_{x}(v)).
\end{align}
Because $h$ is positive definite, $g_{x}(v,v)=0$ if and only if $\mathrm{d}f_{x}(v)=0$.
It follows that $\mathrm{d}f_{x}$ is injective if and only if $g_{x}$ is positive definite. Since $g_{x}$ is positive semidefinite, this is further equivalent to $g_{x}$ being nondegenerate. Since $\dim T_{x}X = n$, injectivity of $\mathrm{d}f_{x}$ is equivalent to $\mathrm{rank}\ \mathrm{d}f_{x} = n$. Finally, because $g_{x}$ is positive semidefinite, it is nondegenerate if and only if $\det g(x) >0$. $\blacksquare$ \\

In applications below, $f=P_{\mathbb{H}}$ and $h=g_{\mathrm{FS}}$. Lemma \ref{lemma 4} distinguishes the fixed complex rank of the projector $P$ from the rank of the differential $\mathrm{d}P_{\mathbb{H};x}$, which controls the local geometry of the projector map. Since $\dim X =4$, the pullback metric is nondegenerate at $x$ precisely when $\mathrm{rank}\ \mathrm{d}P_{\mathbb{H};x} = 4$, that is, when $P_{\mathbb{H}}$ is immersive at $x$. Accordingly, metric degeneracy and failure of immersion are treated as equivalent throughout this work.

\subsection{Proofs of the Main Results}
\begin{theoremAPPENDIX} 
Let $X$ be an oriented four-manifold and suppose that $P$ factors through a quaternionic projector map $P_{\mathbb{H}}: X \rightarrow \mathbb{H}P^{N-1}$. Let $g=P_{\mathbb{H}}^{*}g_{\mathrm{FS}}$ be the pullback Fubini-Study metric, and let $F$ be the associated $\mathrm{SU}(2)$ Berry curvature. Then, for every $x \in X$, 
\begin{align}
& \frac{1}{12}|[\mathrm{Tr}(F\wedge F)]_{1234}(x)| \le \sqrt{\det g(x)}.
\end{align}
At each point $x\in X$, the following two cases occur. \\
\noindent
$\mathrm{(i)}$ $\mathrm{rank}(\mathrm{d}P_{\mathbb{H};x}) < 4$: both sides vanish, and hence the equality holds automatically. \\
\noindent
$\mathrm{(ii)}$ $\mathrm{rank}(\mathrm{d}P_{\mathbb{H};x}) = 4$: the equality holds if and only if the four-dimensional subspace $\mathrm{d}P_{\mathbb{H};x}(T_{x}X)$ is invariant under the canonical quaternionic structure $\mathcal{Q}_{\mathrm{FS}} \subset \mathrm{End}(T\mathbb{H}P^{N-1})$. Equivalently, the rank-three subspace
\begin{align}
& \mathcal{Q}_{x} := (\mathrm{d}P_{\mathbb{H};x})^{-1} \circ \mathcal{Q}_{\mathrm{FS}} \circ \mathrm{d}P_{\mathbb{H};x} \subset \mathrm{End}(T_{x}X)
\end{align}
determines a well-defined $g_{x}$-compatible linear quaternionic structure on $T_{x}X$. 

\end{theoremAPPENDIX}

$\quad Proof.$ 
Fix $x\in X$. Choose a positively oriented basis $(e_{1},e_{2},e_{3},e_{4})$ of $T_{x}X$, and set $v_{\mu}=\mathrm{d}P_{\mathbb{H};x}(e_{\mu})$ for $\mu=1,2,3,4$. Since $g = P^{*}_{\mathbb{H}}g_{\mathrm{FS}}$, we have $g_{\mu\nu}(x) := g_{x}(e_{\mu},e_{\nu}) = g_{\mathrm{FS}}(v_{\mu},v_{\nu})$. 
By Lemma \ref{lemma 3}, in the convention adopted here, the canonical quaternionic fundamental 4-form on $\mathbb{H}P^{N-1}$ is related to the $\mathrm{SU}(2)$ Berry curvature by 
\begin{align}
P^{*}_{\mathbb{H}}\Omega_{\mathrm{FS}} 
& = \frac{1}{6}\sum_{i=1}^{3}\nonumber(P^{*}_{\mathbb{H}}\omega^{i}_{\mathrm{FS}}) \wedge (P^{*}_{\mathbb{H}}\omega^{i}_{\mathrm{FS}}) \nonumber \\
& = \frac{1}{24}\sum_{i=1}^{3}(P^{*}_{\mathbb{H}}\mathcal{F}^{i}) \wedge (P^{*}_{\mathbb{H}}\mathcal{F}^{i}) \nonumber \\
& = \frac{1}{24}\sum_{i=1}^{3} F^{i} \wedge F^{i} 
= -\frac{1}{12}\mathrm{Tr}(F\wedge F),
\end{align}
where $\mathrm{Tr}(T_{i}T_{j})=-\delta_{ij}/2$. The quaternionic Wirtinger inequality applied to $v_{1},\ldots,v_{4}$ gives \cite{Tasaki1985WirtingerInequality,Tasaki1986WirtingerIneqaulity}
\begin{align}
& |\Omega_{\mathrm{FS}}(v_{1},v_{2},v_{3},v_{4})| \le 
\sqrt{\det [g_{\mathrm{FS}}(v_{\mu},v_{\nu})]}.
\end{align}
Using 
\begin{align}
\Omega_{\mathrm{FS}}(v_{1},v_{2},v_{3},v_{4})
& = (P^{*}_{\mathbb{H}}\Omega_{\mathrm{FS}})_{x}(e_{1},e_{2},e_{3},e_{4}) \nonumber \\
& = -\frac{1}{12}[\mathrm{Tr}(F\wedge F)]_{1234}(x),
\end{align}
we obtain
\begin{align}
& \frac{1}{12}|[\mathrm{Tr}(F\wedge F)]_{1234}(x)| \le
\sqrt{\det g(x)}.
\end{align}

Suppose first that $\mathrm{rank}(\mathrm{d}P_{\mathbb{H};x}) < 4$. By Lemma \ref{lemma 4}, $\det g(x)=0$. Moreover, every 4-form vanishes when evaluated on $(v_{1},\ldots,v_{4})$, since these vectors are linearly dependent. By Lemma \ref{lemma 3}, this implies $[\mathrm{Tr}(F\wedge F)]_{1234}(x)=0$. Thus both sides of the inequality vanish, and equality holds automatically. 
Now, suppose that $\mathrm{rank}(\mathrm{d}P_{\mathbb{H};x}) = 4$. Then, 
\begin{align}
& \xi_{x} = \mathrm{d}P_{\mathbb{H};x}(T_{x}X) \subset T_{P_{\mathbb{H}}(x)}\mathbb{H}P^{N-1}
\end{align}
is an oriented real four-plane, and $\mathrm{d}P_{\mathbb{H};x}: T_{x}X \rightarrow \xi_{x}$ is a linear isomorphism. 
By the equality characterization of the quaternionic Wirtinger inequality, equality holds if and only if $\xi_{x}$ is a quaternionic line in $T_{P_{\mathbb{H}}(x)}\mathbb{H}P^{N-1}$ \cite{Tasaki1985WirtingerInequality,Tasaki1986WirtingerIneqaulity}. We now express this condition in terms of the pullback geometry on $T_{x}X$.

Suppose first that equality holds. Then $\xi_{x}$ is a quaternionic line in $T_{P_{\mathbb{H}}(x)}\mathbb{H}P^{N-1}$. Thus, for a local oriented admissible frame $\{I_{\mathrm{FS}}^{i}\}_{i=1}^{3}$ of the canonical quaternionic structure $\mathcal{Q}_{\mathrm{FS}}$, 
\begin{align}
& I_{\mathrm{FS}}^{i}(\xi_{x}) = \xi_{x},\quad i=1,2,3.
\end{align}
Since $\mathrm{d}P_{\mathbb{H};x}$ is a linear isomorphism, we may define 
\begin{align}
& I^{i} = (\mathrm{d}P_{\mathbb{H};x})^{-1} \circ I^{i}_{\mathrm{FS}} \circ \mathrm{d}P_{\mathbb{H};x},\quad i=1,2,3.
\end{align}
Since the endomorphisms $\{I^{i}\}_{i=1}^{3}$ are obtained from $\{I^{i}_{\mathrm{FS}}\}_{i=1}^{3}$, they satisfy the same quaternionic relations. Moreover, for any $v,w \in T_{x}X$,
\begin{align}
g_{x}(I^{i}v,I^{i}w)
& = (P^{*}_{\mathbb{H}}g_{\mathrm{FS}})_{x}(I^{i}v,I^{i}w) \nonumber \\
& = g_{\mathrm{FS}}( \mathrm{d}P_{\mathbb{H};x}(I^{i}v), \mathrm{d}P_{\mathbb{H};x}(I^{i}w) ) \nonumber \\
& = g_{\mathrm{FS}}( I_{\mathrm{FS}}^{i}(\mathrm{d}P_{\mathbb{H};x}v), I_{\mathrm{FS}}^{i}(\mathrm{d}P_{\mathbb{H};x}w) ) \nonumber \\
& = g_{\mathrm{FS}}( \mathrm{d}P_{\mathbb{H};x}v, \mathrm{d}P_{\mathbb{H};x}w) \nonumber \\
& = (P^{*}_{\mathbb{H}}g_{\mathrm{FS}})_{x}(v,w)
= g_{x}(v,w)
\end{align}
Here, we used $g=P^{*}_{\mathbb{H}}g_{\mathrm{FS}}$, $\mathrm{d}P_{\mathbb{H};x} \circ I^{i} = I^{i}_{\mathrm{FS}} \circ \mathrm{d}P_{\mathbb{H};x}$, and the $g_{\mathrm{FS}}$-compatibility of $\{I^{i}_{\mathrm{FS}}\}_{i=1}^{3}$. Therefore,
\begin{align}
& \mathcal{Q}_{x} = \mathrm{span}\{ I^{1},I^{2},I^{3}\} \subset \mathrm{End}(T_{x}X)
\end{align}
determines a well-defined $g_{x}$-compatible linear quaternionic structure on $T_{x}X$.

Conversely, suppose that the pullback subspace $\mathcal{Q}_{x}$ is well-defined. For corresponding oriented admissible frames $\{I^{i}\}_{i=1}^{3}$ and $\{I^{i}_{\mathrm{FS}}\}_{i=1}^{3}$, we then have
\begin{align}
& \mathrm{d}P_{\mathbb{H};x} \circ I^{i} = I^{i}_{\mathrm{FS}} \circ \mathrm{d}P_{\mathbb{H};x},\quad i=1,2,3.
\end{align}
It follows that
\begin{align}
I^{i}_{\mathrm{FS}}(\xi_{x})
& = I^{i}_{\mathrm{FS}}(\mathrm{d}P_{\mathbb{H};x}(T_{x}X))
= \mathrm{d}P_{\mathbb{H};x}(I^{i}(T_{x}X)) \nonumber \\
& = \mathrm{d}P_{\mathbb{H};x}(T_{x}X) = \xi_{x}.
\end{align}
Thus, $\xi_{x}$ is invariant under the canonical quaternionic structure and is therefore a quaternionic line. By the equality characterization, the quaternionic Wirtinger inequality is saturated at $x$. $\blacksquare$

\begin{corollaryAPPENDIX} 
Let the assumptions of Theorem \ref{theorem 1} hold. Suppose that the quaternionic Wirtinger inequality is everywhere saturated and that $\det g(x) > 0$ for every $x \in X$. Then, $P_{\mathbb{H}}$ is an immersion, and 
\begin{align}
& \mathcal{Q} = \bigsqcup_{x\in X} \mathcal{Q}_{x} \subset \mathrm{End}(TX)
\end{align}
defines a smooth rank-three quaternionic structure on $X$ compatible with the metric $g$. Consequently, $(X,g,\mathcal{Q})$ is a quaternionic Hermitian manifold.
\end{corollaryAPPENDIX}

$\quad Proof.$ 
By Lemma \ref{lemma 4}, the assumption $\det g(x)>0$ for every $x\in X$ implies that $P_{\mathbb{H}}$ is an immersion. Since the quaternionic Wirtinger inequality is saturated everywhere, Theorem \ref{theorem 1} implies that 
\begin{align}
& \xi_{x} = \mathrm{d}P_{\mathbb{H};x}(T_{x}X)
\end{align}
is invariant under the canonical quaternionic structure $\mathcal{Q}_{\mathrm{FS}}$ at every $x \in X$. Consequently, the pullback 
\begin{align}
& \mathcal{Q}_{x} = (\mathrm{d}P_{\mathbb{H};x})^{-1} \circ \mathcal{Q}_{\mathrm{FS}} \circ \mathrm{d}P_{\mathbb{H};x} 
\subset \mathrm{End}(T_{x}X)
\end{align}
determines a well-defined $g_{x}$-compatible linear quaternionic structure on $T_{x}X$.

Since $P_{\mathbb{H}}$ is an immersion, each map $\mathrm{d}P_{\mathbb{H};x}: T_{x}X \rightarrow \mathrm{d}P_{\mathbb{H};x}(T_{x}X)$ is a linear isomorphism. Moreover, because $\mathrm{d}P_{\mathbb{H}}$ depends smoothly on $x$ and has constant rank four, its inverse on the image also varies smoothly with $x$. Together with the smoothness of the canonical quaternionic structure $\mathcal{Q}_{\mathrm{FS}}$, this implies that the family $\{\mathcal{Q}_{x}\}_{x\in X}$ varies smoothly. Therefore, 
\begin{align}
& \mathcal{Q} = \bigsqcup_{x\in X}\mathcal{Q}_{x} \subset \mathrm{End}(TX)
\end{align}
defines a smooth rank-three almost quaternionic structure on $X$. 
Since $X$ is four-dimensional, $\mathcal{Q}$ is automatically a quaternionic structure in the sense of Definition \ref{definition 3}. Since each $\mathcal{Q}_{x}$ is $g_{x}$-compatible by construction, $\mathcal{Q}$ is compatible with $g$. Consequently, by Definition \ref{definition 4}, $(X,g,\mathcal{Q})$ is a quaternionic Hermitian manifold. $\blacksquare$

\begin{corollaryAPPENDIX}
Let the assumptions of Theorem \ref{theorem 1} hold. If $X$ is closed, then
\begin{align} \label{lower bound}
& \frac{2\pi^{2}}{3}|\mathcal{C}_{2}| \le \int_{X}\sqrt{\det g(x)}\ \mathrm{d}^{4}x,
\end{align}
where 
\begin{align}
& \mathcal{C}_{2} = -\frac{1}{8\pi^{2}} \int_{X} \mathrm{Tr}(F\wedge F) \in \mathbb{Z}
\end{align}
is the second Chern number of the occupied complex rank-two bundle defined by $P$.
Equality holds if and only if the quaternionic Wirtinger inequality is saturated at every point of $X$ and $\mathrm{Tr}(F\wedge F)$ does not change sign with respect to the orientation of $X$.
\end{corollaryAPPENDIX}

$\quad Proof.$ 
By the definition of the second Chern number and the triangle inequality,
\begin{align} \label{triangle}
|\mathcal{C}_{2}| 
& = \frac{1}{8\pi^{2}} \biggr|\int_{X} \mathrm{Tr}(F\wedge F)\biggr| \nonumber \\
& = \frac{1}{8\pi^{2}} \biggr|\int_{X} [\mathrm{Tr}(F\wedge F)]_{1234}(x)\mathrm{d}^{4}x\biggr| \nonumber \\
& \le \frac{1}{8\pi^{2}} \int_{X} \Bigr|[\mathrm{Tr}(F\wedge F)]_{1234}(x)\Bigr| \mathrm{d}^{4}x \nonumber \\
& \le \frac{3}{2\pi^{2}} \int_{X} \sqrt{\det g(x)} \mathrm{d}^{4}x.
\end{align}
Multiplying both sides by $2\pi^{2}/3$ yields
\begin{align}
\frac{2\pi^{2}}{3} |\mathcal{C}_{2}| 
& \le \int_{X} \sqrt{\det g(x)} \mathrm{d}^{4}x. 
\end{align}
Equality in the first inequality in Eq. (\ref{triangle}) holds if and only if $\mathrm{Tr}(F\wedge F)$ is either nonnegative or nonpositive throughout $X$, with respect to the chosen orientation. Equality in the second holds if and only if the quaternionic Wirtinger inequality is saturated everywhere. Hence, equality in Eq. (\ref{lower bound}) holds precisely under the stated conditions. $\blacksquare$

\begin{theoremAPPENDIX} 
Consider a gapped four-dimensional four-band Bloch Hamiltonian $h(\bm{\mathrm{k}})$ at half-filling with a fixed antiunitary operator $\mathcal{J}:\mathbb{C}^{4}\rightarrow\mathbb{C}^{4}$ satisfying
\begin{align}
& \mathcal{J}^{2}=-\bm{1}_{4}, \quad 
\mathcal{J} h(\bm{\mathrm{k}})\mathcal{J}^{-1}=h(\bm{\mathrm{k}})
\end{align}
for every $\bm{\mathrm{k}} \in T^{4}$. 
Then, there must exist a point $\bm{\mathrm{k}}_{*} \in T^{4}$ at which $\det g(\bm{\mathrm{k}}_{*}) = 0$. Equivalently, the quaternionic projector
\begin{align}
& P_{\mathbb{H}}: T^{4} \rightarrow \mathbb{H}P^{1} \cong S^{4}
\end{align}
cannot be an immersion.
\end{theoremAPPENDIX}

$\quad Proof.$ 
Since $\mathcal{J}^{2}=-\bm{1}_{4}$ and the fixed operator $\mathcal{J}$ commutes with $h(\bm{\mathrm{k}})$ for every $\bm{\mathrm{k}} \in T^{4}$, the occupied complex two-plane at each momentum is $\mathcal{J}$-invariant and therefore defines a quaternionic line. Hence, the complex projector $P:T^{4} \rightarrow \mathrm{Gr}_{2,4}^{\mathbb{C}}$ takes values in the $\mathcal{J}$-invariant locus $(\mathrm{Gr}_{2,4}^{\mathbb{C}})^{\mathcal{J}}$. By Lemma \ref{lemma 2}, it factors uniquely as $P=\eta_{1,2}\circ P_{\mathbb{H}}$ where
\begin{align}
& P_{\mathbb{H}}: T^{4} \rightarrow \mathbb{H}P^{1} \cong S^{4}.
\end{align}

Suppose that $P_{\mathbb{H}}$ were an immersion. Since $T^{4}$ and $S^{4}$ have the same dimension, $P_{\mathbb{H}}$ would be a local diffeomorphism. Moreover, since $T^{4}$ is compact and $S^{4}$ is Hausdorff, $P_{\mathbb{H}}$ is proper. Propositions 4.8(b), A.53(a), and 4.46 of Ref.~\cite{LeeSmoothManifolds} therefore imply that $P_{\mathbb{H}}$ would be a smooth covering map. 
By Proposition 1.31 of Ref.~\cite{HatcherAlgebraicTopology}, it would induce an injective homomorphism 
\begin{align}
& (P_{\mathbb{H}})_{*}: \pi_{1}(T^{4}) \rightarrow \pi_{1}(S^{4}).
\end{align}
However,
\begin{align}
& \pi_{1}(T^{4}) \cong \mathbb{Z}^{4}, \quad \pi_{1}(S^{4}) = 0,
\end{align}
so no such injective homomorphism exists. Therefore, $P_{\mathbb{H}}$ cannot be an immersion. 
By Lemma \ref{lemma 4}, there must exist $\bm{\mathrm{k}}_{*} \in T^{4}$ such that 
\begin{align}
& \det g(\bm{\mathrm{k}}_{*}) = 0.\quad \blacksquare
\end{align}

\begin{theoremAPPENDIX} 
Consider a gapped four-dimensional four-band Bloch Hamiltonian $h(\bm{\mathrm{k}})$ at half-filling with a fixed antiunitary operator $\mathcal{J}:\mathbb{C}^{4}\rightarrow\mathbb{C}^{4}$ satisfying
\begin{align}
& \mathcal{J}^{2}=-\bm{1}_{4}, \quad 
\mathcal{J} h(\bm{\mathrm{k}})\mathcal{J}^{-1}=h(\bm{\mathrm{k}})
\end{align}
for every $\bm{\mathrm{k}} \in T^{4}$. 
Then, the quaternionic Wirtinger inequality is everywhere saturated 
\begin{align}
& \frac{1}{12}|[\mathrm{Tr}(F\wedge F)]_{1234}(\bm{\mathrm{k}})| = \sqrt{\det g(\bm{\mathrm{k}})}.
\end{align}
\end{theoremAPPENDIX}

$\quad Proof.$ 
As established in the proof of Theorem \ref{theorem 2}, the complex projector $P:T^{4} \rightarrow \mathrm{Gr}_{2,4}^{\mathbb{C}}$ factors uniquely as $P=\eta_{1,2}\circ P_{\mathbb{H}}$ where
\begin{align}
& P_{\mathbb{H}}: T^{4} \rightarrow \mathbb{H}P^{1} \cong S^{4}.
\end{align}
Fix $\bm{\mathrm{k}} \in T^{4}$. Suppose first that $P_{\mathbb{H}}$ is immersive at $\bm{\mathrm{k}}$. Since both $T^{4}$ and $\mathbb{H}P^{1}$ have real dimension four, $\mathrm{d}P_{\mathbb{H};\bm{\mathrm{k}}}$ is an isomorphism and
\begin{align}
& \mathrm{d}P_{\mathbb{H};\bm{\mathrm{k}}} (T_{\bm{\mathrm{k}}}T^{4}) 
= T_{P_{\mathbb{H}}(\bm{\mathrm{k}})}\mathbb{H}P^{1}.
\end{align}
The latter is a quaternionic vector space of quaternionic dimension one and therefore invariant under the canonical quaternionic structure. Hence, the equality condition is automatically satisfied at $\bm{\mathrm{k}}$:
\begin{align}
& \frac{1}{12} |\left[ \mathrm{Tr}(F \wedge F) \right]_{1234}(\bm{\mathrm{k}})| = \sqrt{\det g(\bm{\mathrm{k}})}.
\end{align}
Suppose instead that $P_{\mathbb{H}}$ is not immersive at $\bm{\mathrm{k}}$. By Lemma \ref{lemma 4}, $\det g(\bm{\mathrm{k}})=0$. Moreover, $\mathrm{rank}\ \mathrm{d}P_{\mathbb{H};\bm{\mathrm{k}}} < 4$, so the pullback of any 4-form on $\mathbb{H}P^{1}$ vanishes at $\bm{\mathrm{k}}$. Since $\mathrm{Tr}(F\wedge F)$ is the pullback of the corresponding 4-form on $\mathbb{H}P^{1}$, it follows that 
\begin{align}
& [\mathrm{Tr}(F \wedge F)]_{1234}(\bm{\mathrm{k}})=0.
\end{align}
Therefore, the equality holds at every $\bm{\mathrm{k}} \in T^{4}$. $\blacksquare$

\begin{theoremAPPENDIX} 
For $N\ge2$, consider a smooth quaternionic projector map $P_{\mathbb{H}}: T^{4} \rightarrow \mathbb{H}P^{N-1}$ associated with a quaternionic line, and write $g=P_{\mathbb{H}}^{*}g_{\mathrm{FS}}$ for the corresponding quantum metric. Suppose that the quaternionic Wirtinger inequality is saturated at every $\bm{\mathrm{k}} \in T^{4}$. 
Then, there must exist a point $\bm{\mathrm{k}}_{*} \in T^{4}$ at which $\det g(\bm{\mathrm{k}}_{*})=0$. Equivalently, $P_{\mathbb{H}}$ cannot be an immersion. 
\end{theoremAPPENDIX}

$\quad Proof.$ 
Suppose, to the contrary, that the quantum metric is nondegenerate everywhere, $\det g(\bm{\mathrm{k}}) > 0$ for all $\bm{\mathrm{k}} \in T^{4}$. By Lemma \ref{lemma 4}, $P_{\mathbb{H}}$ is then an immersion on all of $T^{4}$. Since $T^{4}$ is connected, Proposition S7 of Ref.~\cite{Yang2026QKgeometryTRS} applied with $U=T^{4}$, together with the assumed saturation, implies that the entire image of $P_{\mathbb{H}}$ is contained in a fixed totally geodesic quaternionic projective line,
\begin{align}
& P_{\mathbb{H}}(T^{4}) \subset \mathbb{H}P^{1} \subset \mathbb{H}P^{N-1}.
\end{align}
Hence, $P_{\mathbb{H}}$ can be regarded as a smooth immersion
\begin{align}
& P_{\mathbb{H}}: T^{4} \rightarrow \mathbb{H}P^{1} \cong S^{4}.
\end{align}
This contradicts the topological argument used in the proof of  Theorem \ref{theorem 2}, according to which no smooth immersion from $T^{4}$ into $\mathbb{H}P^{1}\cong S^{4}$ exists. Therefore, there must exist at least one $\bm{\mathrm{k}}_{*} \in T^{4}$ such that $\det g(\bm{\mathrm{k}}_{*})=0$. Equivalently, by Lemma \ref{lemma 4}, $P_{\mathbb{H}}$ is not an immersion. $\blacksquare$

\section{Quaternionic Coherent-State Construction}
\label{appendix C}
\subsection{General Perelomov Construction}
Let $G$ be an arbitrary Lie group and $T$ be an irreducible unitary representation on a complex Hilbert space $\mathcal{H}_{\mathbb{C}}$, and let $|\psi_{e}\rangle$ be a fixed normalized reference vector in $\mathcal{H}_{\mathbb{C}}$. Consider the set of vectors 
\begin{align}
& \{ |\psi_{g}\rangle \} = \{ |\psi_{g}\rangle = T(g)|\psi_{e}\rangle \in \mathcal{H}_{\mathbb{C}} ; g \in G \}. 
\end{align}
If the two states $|\psi_{g}\rangle$ and $|\psi_{g'}\rangle$ are in the same complex ray, there exists a phase factor $e^{i\alpha}$ satisfying $|\psi_{g}\rangle = e^{i\alpha}|\psi_{g'}\rangle$. 
Let us define the stabilizer $H$ by
\begin{align}
& H = \{ h \in G ;  T(h)|\psi_{e}\rangle=e^{i\alpha(h)}|\psi_{e}\rangle \}.
\end{align}
Since all elements of a coset $gH$ determine the same complex ray, the quotient space $X=G/H$ is a homogeneous parameter space of coherent-state rays. We denote the corresponding ray by 
\begin{align}
& [|\psi_{x}\rangle] = [T(g_{x})|\psi_{e}\rangle] \text{ for each } x = g_{x}H \in X.
\end{align}

The Perelomov construction reviewed above is formulated in an ordinary complex Hilbert space, in which two vectors that differ by a $\mathrm{U}(1)$ phase represent the same complex ray. However, this does not require the symmetry group $G$ or its stabilizer to be Abelian. In the minimal-charge $S^{4}$ Landau level, the physical state is a $\mathcal{J}$-invariant complex two-plane, or equivalently a quaternionic line. Choosing a complex ray within this quaternionic line retains an additional $\mathrm{Sp}(1)/\mathrm{U}(1) \cong S^{2}$ degree of freedom \cite{Zhang2001FourQHE,Polchinski2003QHEofR4}. This distinction is reflected in the quotient descriptions
\begin{align}
& \mathbb{C}P^{2N-1} \cong \frac{S^{4N-1}}{\mathrm{U}(1)}, \quad
\mathbb{H}P^{N-1} \cong \frac{S^{4N-1}}{\mathrm{Sp}(1)}.
\end{align}
We therefore use the natural quaternionic ray formulation, which quotients out this additional degree of freedom.

Let $T$ now be a quaternionic-unitary representation on a quaternionic Hilbert space $\mathcal{H}_{\mathbb{H}}$, and let $|\psi_{e}\rangle$ be a fixed normalized reference vector in $\mathcal{H}_{\mathbb{H}}$. 
Consider two normalized quaternionic representatives
\begin{align}
& |\psi_{x}\rangle =T(g_{x})|\psi_{e}\rangle, \quad |\psi_{x}'\rangle=T(g_{x}h)|\psi_{e}\rangle,
\end{align}
associated with $g_{x}$ and $g_{x}h$, which represent the same coset $x = g_{x}H \in G/H$. Adopting the right multiplication convention, the stabilizer satisfies 
\begin{align}
& T(h)|\psi_{e}\rangle = |\psi_{e}\rangle q_{h}, \quad q_{h} \in \mathrm{Sp}(1),
\end{align}
for all $h \in H$. Hence,
\begin{align}
& |\psi_{x}'\rangle
= T(g_{x})T(h)|\psi_{e}\rangle
= T(g_{x}) |\psi_{e}\rangle q_{h}
= |\psi_{x}\rangle q_{h}.
\end{align}
Since $q_{h}q_{h}^{\dagger}=1$, we find
\begin{align}
& |\psi_{x}'\rangle \langle \psi_{x}'|
= |\psi_{x}\rangle (q_{h} q_{h}^{\dagger}) \langle \psi_{x}|
= |\psi_{x}\rangle \langle\psi_{x}|.
\end{align}
Therefore, the quaternionic rank-one projector 
\begin{align}
& P_{\mathbb{H}}(x)=|\psi_{x}\rangle \langle \psi_{x}|
\end{align}
depends only on the coset $x \in G/H$, not on the choice of its representative. Under the complex realization, the quaternionic representative $|\psi_{x}\rangle$ becomes a $\mathcal{J}$-related complex frame $\Psi_{x} \in \mathbb{C}^{2N\times2}$. By Lemma \ref{lemma 2}, the corresponding complex projector is  
\begin{align} \label{projector btw complex and quaternion}
& P(x) = \eta_{1,N}(P_{\mathbb{H}}(x)) = \Psi_{x}\Psi_{x}^{\dagger},
\end{align}
which has complex rank-two and is $\mathcal{J}$-invariant.

\subsection{Coset Parameterization of the Four-Sphere}
For the Lie group $G=\mathrm{Sp}(2)$ and the gauge group $\Lambda = \mathrm{Sp}(1) \cong \mathrm{SU}(2)$, the stabilizer $H$ is given by \cite{Perelomov1986BookCoherentStates,Perelomov2002CoherentState}
\begin{align}
& H = \{h\in \mathrm{Sp}(2); T(h)|\psi_{e}\rangle = |\psi_{e}\rangle q \text{ for some } q \in \Lambda \}.
\end{align}
To determine $H$ explicitly, we write
\begin{align} \label{the condition of G}
& |\psi_{e}\rangle  = \begin{pmatrix}
1 \\ 0
\end{pmatrix} \text{ and }
 T(h) = \begin{pmatrix}
a & b \\ c & d
\end{pmatrix},
\end{align}
for $a,b,c,d\in \mathbb{H}$. This condition implies that $\begin{pmatrix}a & c\end{pmatrix}^{T} = \begin{pmatrix}q & 0\end{pmatrix}^{T}$, so $a=q$ and $c=0$. Since $T(h)^{\dagger}T(h)=I$, we obtain $b=0$ and $\bar{d}d=1$, which implies $d \in \mathrm{Sp}(1)$. As a result, we find
\begin{align}
& H = \mathrm{Sp}(1) \times \mathrm{Sp}(1),
\end{align}
and the parameter space is given by
\begin{align}
& X = G/H = \mathrm{Sp}(2) / \mathrm{Sp}(1) \times \mathrm{Sp}(1) \cong S^{4} \cong \mathbb{H}P^{1}.
\end{align}

Applying the complex realization in Eq. (\ref{projector btw complex and quaternion}) to $\mathbb{H}^{2}$, the quaternionic reference state $|\psi_{e}\rangle$ is represented by the $\mathcal{J}$-related complex frame \cite{Kuratsuji1990QuaternionicPerelomovCoherentState}:
\begin{align}
& \Psi_{e} = \begin{pmatrix} \bm{1}_{2} \\ 0 \end{pmatrix} \in \mathbb{C}^{4\times2}.
\end{align}
Let $x \in S^{4}$ be represented by 
\begin{align}
& x = (
\sin \theta_{x}n_{1}, \sin\theta_{x}n_{2}, 
\sin \theta_{x}n_{3}, \sin\theta_{x}n_{4},
\cos\theta_{x})
\end{align}
with $\sum_{\mu=1}^{4} n_{\mu}^{2} = 1$. 
We associate with $n=(n_{1},n_{2},n_{3},n_{4}) \in S^{3}$ the unit quaternion
\begin{align}
& \lambda_{x} = n_{4}\bm{1}_{2} - i\sum_{i=1}^{3} n_{i}\sigma_{i} \in \mathrm{SU}(2).
\end{align}
On a local patch, a convenient coset representative $g_{x} \in \mathrm{Sp}(2)$ that maps the north pole to $x$ and commutes with $\mathcal{J}=-(\bm{1}_{2}\otimes i\sigma_{2})\mathcal{K}$ is 
\begin{align}
& g_{x} = \begin{pmatrix}
\cos\tfrac{\theta_{x}}{2} \bm{1}_{2} & -\sin\tfrac{\theta_{x}}{2} \lambda_{x}^{\dagger} \\
\sin\tfrac{\theta_{x}}{2} \lambda_{x} & \cos\tfrac{\theta_{x}}{2} \bm{1}_{2}
\end{pmatrix}.
\end{align}
The Perelomov coherent-state frame associated with $x$ is obtained by acting on the reference frame: 
\begin{align} \label{coherent state of S4}
& \Psi_{x} = T(g_{x})\Psi_{e} = 
\begin{pmatrix}
\cos\frac{\theta_{x}}{2}\bm{1}_{2} \\
\sin\frac{\theta_{x}}{2}\lambda_{x}
\end{pmatrix}.
\end{align}
Using the relations
\begin{align}
& \cos\frac{\theta_{x}}{2} = \sqrt{\frac{1+x_{5}}{2}}, \quad
\sin\frac{\theta_{x}}{2} \lambda_{x} = \frac{x_{4}\bm{1}_{2}-ix_{i}\sigma_{i}}{\sqrt{2(1+x_{5})}},
\end{align}
the coherent-state frame in Eq. (\ref{coherent state of S4}) can be written directly in terms of the embedding coordinates $x_{a}$ on $S^{4}$ as
\begin{align}
& \Psi_{x}
= \begin{pmatrix} \sqrt{\frac{1+x_{5}}{2}}\bm{1}_{2} \\ \sqrt{\frac{1}{2(1+x_{5})}}(x_{4}\bm{1}_{2}-ix_{i}\sigma_{i}) \end{pmatrix}.
\end{align}

The two columns of $\Psi_{x}$ form an orthonormal basis of a quaternionic line, with $\Psi_{x}^{\dagger}\Psi_{x}=\bm{1}_{2}$. 
Although $\Psi_{x}$ is defined only locally, the associated complex rank-two projector $P(x) = \Psi_{x}\Psi_{x}^{\dagger}$ is independent of the choice of local frame and is therefore globally defined:
\begin{align}
& P(x) 
= \begin{pmatrix}
\cos^{2}\tfrac{\theta_{x}}{2} \bm{1}_{2} & \cos\tfrac{\theta_{x}}{2} \sin\tfrac{\theta_{x}}{2} \lambda_{x}^{\dagger} \\
\cos\tfrac{\theta_{x}}{2} \sin\tfrac{\theta_{x}}{2} \lambda_{x} & 
\sin^{2}\tfrac{\theta_{x}}{2} \bm{1}_{2}
\end{pmatrix}.
\end{align}
In terms of the Gamma matrices defined in Eq. (\ref{gamma representation}), the projector takes the compact form
\cite{Hasebe2020YangMonopoleAndBPSTInstanton}:
\begin{align}
& P(x) = \frac{1}{2}(\bm{1}_{4} + x_{a}\Gamma_{a}).
\end{align}

In the construction of Ref. \cite{Zhang2001FourQHE}, $x\in S^{4}$ labels a point on the real-space four-sphere. The real-space four-sphere and the parameter-space four-sphere are both realized as the same homogeneous space $S^{4}\cong\mathbb{H}P^{1}$. Their common $\mathrm{Sp}(2)$ action therefore provides a natural equivariant identification arising from the shared homogeneous-space realization. An analogous identification holds on $S^{2}$. If we choose the reference state $|\psi_{e}\rangle=(1,0)^{T}\in \mathbb{C}^{2}$ and the coset representative $g_{x} = e^{i\phi_{x}\sigma_{3}/2}e^{-i\theta_{x}\sigma_{2}/2}$ \cite{Arecchi1972AtomicCoherentStateinQuantumOptics}, the corresponding coherent state is
\begin{align}
& |\psi_{x}\rangle = T(g_{x})|\psi_{e}\rangle = \begin{pmatrix}
\cos\tfrac{\theta_{x}}{2}e^{i\phi_{x}/2} \\
\sin\tfrac{\theta_{x}}{2}e^{-i\phi_{x}/2}
\end{pmatrix} 
= \begin{pmatrix} \alpha \\ \beta \end{pmatrix}.
\end{align}
For the magnetic monopole charge $I$, Haldane expressed the LLL wavefunction localized around the point $x\in S^{2}$ by 
\begin{align}
& \psi_{x}^{(I)}(u,v) = (\alpha^{*}u+\beta^{*}v)^{2I},
\end{align}
where $u= \cos\tfrac{\theta}{2}e^{i\phi/2}$, $v=\sin\tfrac{\theta}{2}e^{-i\phi/2}$. Here, $[u:v]\in \mathbb{C}P^{1}$ denotes the real-space coordinate, whereas $[\alpha:\beta] \in \mathbb{C}P^{1}$ denotes the parameter-space coordinate and labels the localization center $x$ \cite{Haldane1983QHE}. Thus, the real-space two-sphere and the parameter-space two-sphere are equivariantly identified, exactly as in the $S^{4}$ construction.

Unlike the spherical construction above, real space and the BZ of a four-dimensional periodic lattice form a dual pair. The Bloch construction provides no analogous common transitive group action that canonically identifies their points. The identifications found on $S^{2}$ and $S^{4}$ therefore arise from their shared coset realizations, rather than from a generic relation between real and parameter spaces.

\section{Explicit Quaternionic Hermitian Band Geometry on the Four-Sphere} 
\label{appendix D}
We use the complex $\mathrm{SU}(2)$ matrix representation throughout this appendix. Under the canonical identification of the tautological quaternionic line bundle with the restriction of the complex rank-two tautological bundle, $\eta: \mathbb{H}P^{N-1} \rightarrow (\mathrm{Gr}_{2,2N}^{\mathbb{C}})^{\mathcal{J}}$ is an isometry in the complex rank-two normalization. Moreover, the pullback along $\eta$ of the universal $\mathrm{U}(2)$ connection preserves the quaternionic structure and hence reduces to the canonical $\mathrm{SU}(2)\cong\mathrm{Sp}(1)$ connection. Hence,
\begin{align}
& g_{\mathrm{FS}} = \eta^{*}g_{\mathrm{FS}}^{\mathbb{C}}, \quad
\mathcal{F}_{\mathbb{H}} = \eta^{*}\mathcal{F}_{\mathbb{C}}.
\end{align}
Since $P = \eta \circ P_{\mathbb{H}}$, the naturality of pullbacks gives
\begin{align}
P^{*}g_{\mathrm{FS}}^{\mathbb{C}}
& = (\eta \circ P_{\mathbb{H}})^{*}g_{\mathrm{FS}}^{\mathbb{C}}
= P_{\mathbb{H}}^{*} ( \eta^{*}g_{\mathrm{FS}}^{\mathbb{C}})
= P_{\mathbb{H}}^{*} g_{\mathrm{FS}}, \nonumber \\
P^{*}\mathcal{F}_{\mathbb{C}}
& = (\eta \circ P_{\mathbb{H}})^{*}\mathcal{F}_{\mathbb{C}}
= P_{\mathbb{H}}^{*} ( \eta^{*}\mathcal{F}_{\mathbb{C}})
= P_{\mathbb{H}}^{*} \mathcal{F}_{\mathbb{H}}.
\end{align}
Thus, the complex-projector calculations below yield precisely the pullback metric and Berry curvature entering Theorem \ref{theorem 1}.

\subsection{$\mathrm{SU}(2)$ Berry Connection and Curvature}
The local coherent-state frame $\Psi_{x}$ introduced in Sec. \ref{section 2} represents the globally defined complex rank-two projector $P=\eta \circ P_{\mathbb{H}}$:
\begin{align}
& P: S^{4} \rightarrow (\mathrm{Gr}_{2,4}^{\mathbb{C}})^{\mathcal{J}}, \quad x \mapsto \Psi_{x}\Psi_{x}^{\dagger}.
\end{align}
The non-Abelian Berry connection and its curvature are
\begin{align}
& A = \Psi_{x}^{\dagger}\mathrm{d}\Psi_{x}, \quad 
F = P^{*}\mathcal{F}_{\mathbb{C}} =\mathrm{d}A + A \wedge A.
\end{align}
Under a gauge transformation $g:U \rightarrow \mathrm{SU}(2)$, they transform as
\begin{align}
& A \mapsto g^{-1}Ag + g^{-1}\mathrm{d}g, \quad F \mapsto g^{-1}Fg.
\end{align}
Thus, although $F$ transforms covariantly rather than remaining gauge invariant, $\mathrm{Tr}\ F$ and $\mathrm{Tr}(F \wedge F)$ are gauge invariant.

For convenience, $i$ ranges over $\{1,2,3\}$, $\mu$ ranges over $\{1,2,3,4\}$, and $a$ ranges over $\{1,2,3,4,5\}$. For $y=(y_{1},y_{2},y_{3},y_{4}) \in \mathbb{R}^{4}$, the inverse stereographic projection onto $S^{4} \backslash \{\mathrm{SP}\}$ is given by \cite{Levay2004QuaternionicGeometry}:
\begin{align}
& x_{\mu}=\frac{2y_{\mu}}{1+y^{2}}, \quad x_{5} = \frac{1-y^{2}}{1+y^{2}}.
\end{align}
The origin $y=0$ corresponds to the north pole, whereas the omitted south pole is approached as $|y|\rightarrow \infty$. In these coordinates, the coherent-state frame becomes
\begin{align}
\Psi_{x}
& = \frac{1}{\sqrt{1+y^{2}}} \begin{pmatrix}
\bm{1}_{2} \\
y_{4}\bm{1}_{2} - iy_{i}\sigma_{i}
\end{pmatrix}
= N^{-1/2} \begin{pmatrix}
\bm{1}_{2} \\
\lambda_{x}
\end{pmatrix},
\end{align}
where $N = 1+y^{2}$, $\lambda_{x}= y_{4}\bm{1}_{2} - iy_{i}\sigma_{i}$. Then, the Berry connection is given by
\begin{align}
& A = A^{i}T_{i} = \left(\frac{2\eta_{\mu\nu}^{i}y_{\nu}\mathrm{d}y_{\mu}}{1+y^{2}} \right) T_{i}
\end{align}
where $T_{i}=-i\sigma_{i}/2$ and $\eta_{\mu\nu}^{i}$ are self-dual 't Hooft symbols defined as $\eta^{i}_{jk} = \epsilon_{ijk}, \eta^{i}_{j4} = \delta_{ij}, \eta^{i}_{4k} = -\delta_{ik}$. Writing $F=F^{i}T_{i}$, its components are
\begin{align}
& F^{i} = \mathrm{d}A^{i} + \frac{1}{2}\epsilon_{ijk} A^{j} \wedge A^{k}
\end{align}
with $[T_{i},T_{j}] = \epsilon_{ijk}T_{k}$. The exterior-derivative and quadratic contributions to $F^{i}$ are, respectively,
\begin{align} \label{dA}
\mathrm{d}A^{i}
& = \frac{4}{N^{2}} \eta^{i}_{\mu\nu} y_{\rho} y_{\nu} \mathrm{d}y_{\rho} \wedge \mathrm{d}y_{\mu} + \frac{2}{N} \eta^{i}_{\mu\nu}\mathrm{d}y_{\nu} \wedge \mathrm{d}y_{\mu},
\end{align}
and
\begin{align} \label{AwedgeA}
\frac{1}{2} \epsilon_{ijk} A^{j} \wedge A^{k}
& = \frac{2}{N^{2}} \eta^{i}_{\mu\nu} 
y^{2} \mathrm{d}y_{\mu} \wedge \mathrm{d}y_{\nu} \nonumber \\
& + \frac{4}{N^{2}} \eta^{i}_{\mu\nu} 
y_{\rho}y_{\nu} \mathrm{d}y_{\rho} \wedge \mathrm{d}y_{\mu}.
\end{align}
Combining Eqs. (\ref{dA}) and (\ref{AwedgeA}), the terms proportional to $y_{\rho}y_{\nu}$ cancel, yielding
\begin{align}
& F^{i}_{\mu\nu} = - \frac{4\eta^{i}_{\mu\nu}}{(1+y^{2})^{2}},
\end{align}
and hence
\begin{align} \label{TrFwedgeF}
\mathrm{Tr}(F \wedge F)
& = -\frac{1}{2}\sum_{i=1}^{3} F^{i}\wedge F^{i} 
= -\frac{48}{(1+y^{2})^{4}} \mathrm{d}^{4}y,
\end{align}
where $\mathrm{d}^{4}y = \mathrm{d}y_{1}\wedge\mathrm{d}y_{2}\wedge\mathrm{d}y_{3}\wedge\mathrm{d}y_{4}$.

We next compute the Chern classes and the resulting second Chern number. Since each $T_{i}$ is traceless, the first Chern class vanishes:
\begin{align}
& c_{1}(E) = \frac{i}{2\pi} \mathrm{Tr}\ F
= \frac{i}{2\pi} F^{i}\ \mathrm{Tr}\ T_{i} = 0
\end{align}
The second Chern class is 
\begin{align}
& c_{2}(E) = -\frac{1}{8\pi^{2}} \mathrm{Tr}(F \wedge F)
= \frac{6}{\pi^{2}} \frac{\mathrm{d}^{4}y}{(1+y^{2})^{4}}.
\end{align}
Using this stereographic chart, the second Chern number is
\begin{align}
\mathcal{C}_{2}
& = \int_{S^{4}} c_{2}(E)
= \frac{6}{\pi^{2}} \int_{\mathbb{R}^{4}} \frac{\mathrm{d}^{4}y}{(1+y^{2})^{4}} = 1.
\end{align}
Thus, the first Chern class vanishes, while the second Chern number is $\mathcal{C}_{2}=1$.

\subsection{Quantum Metric and Saturation of the Geometric Bounds}
The non-Abelian quantum metric is defined by
\begin{align}
& g = P^{*}g_{\mathrm{FS}}^{\mathbb{C}} = \mathrm{Tr}( P\mathrm{d}P\mathrm{d}P ).
\end{align}
A direct calculation gives the nondegenerate quantum metric
\begin{align}
g & = \frac{\mathrm{Tr}(\mathrm{d}\lambda_{x}^{\dagger}\mathrm{d}\lambda_{x})}{(1+y^{2})^{2}}
= \frac{2\sum_{\mu=1}^{4}\mathrm{d}y_{\mu}^{2}}{(1+y^{2})^{2}},
\end{align}
or equivalently,
\begin{align} \label{explicit metric of S4}
& g_{\mu\nu} = \frac{2\delta_{\mu\nu}}{(1+y^{2})^{2}}.
\end{align}

Combining Eqs. (\ref{TrFwedgeF}) and (\ref{explicit metric of S4}), we obtain pointwise saturation on the stereographic chart:
\begin{align} \label{S4 saturation}
& \frac{1}{12}| [\mathrm{Tr} (F\wedge F)]_{1234}| = \sqrt{\det g} = \frac{4}{(1+y^{2})^{4}}>0.
\end{align}
The metric in Eq. (\ref{explicit metric of S4}) is one half of the standard round metric and hence extends smoothly and nondegenerately across the omitted south pole. Since both sides of Eq. (\ref{S4 saturation}) are globally defined smooth densities, the equality also extends across the south pole, yielding everywhere nondegenerate saturation. This verifies the hypotheses of Corollary \ref{corollary 1}. Equation (\ref{TrFwedgeF}) further shows that the second Chern density has a fixed sign, thereby completing the verification of the equality conditions of Corollary \ref{corollary 2}. The resulting geometric and global consequences are summarized in Sec. \ref{section 3}.

\section{Metric Degeneracy in the Four-Dimensional Lattice Dirac Model}
\label{appendix E}
We consider the four-dimensional lattice Dirac Hamiltonian introduced in Eq. (\ref{lattice dirac model Hamiltonian}), with the gamma-matrix representation given in Eq. (\ref{gamma representation}). For $c \ne 0$, we assume $m/c \notin \{-4,-2,0,2,4\}$, so that $d(\bm{\mathrm{k}}) \ne 0$ throughout the BZ and the negative-energy eigenspace defines a smooth complex rank-two spectral projector.
We next examine the action of $\mathcal{J}$ on the gamma matrices. For $a=1,2,3$,
\begin{align}
\mathcal{J}\Gamma^{a}\mathcal{J}^{-1}
& = (\bm{1}_{2} \otimes i\sigma_{2})(-\tau_{2}\otimes\sigma_{a})^{*}(\bm{1}_{2} \otimes -i\sigma_{2})
= \Gamma^{a}.
\end{align}
For the remaining two matrices,
\begin{align}
\mathcal{J}\Gamma^{4,5}J^{-1}
= (\bm{1}_{2} \otimes i\sigma_{2})(\tau_{1,3} \otimes \bm{1}_{2})(\bm{1}_{2} \otimes -i\sigma_{2})
= \Gamma^{4,5}.
\end{align}
Therefore, all five gamma matrices are $\mathcal{J}$-invariant:
\begin{align} \label{commutation gamma}
& \mathcal{J}\Gamma^{a}\mathcal{J}^{-1} = \Gamma^{a} \text{ for } a=1,2,3,4,5.
\end{align}
Since all $d_{a}(\bm{\mathrm{k}})$ are real, Eq. (\ref{commutation gamma}) gives, for all $\bm{\mathrm{k}} \in T^{4}$,
\begin{align}
\mathcal{J}h(\bm{\mathrm{k}})\mathcal{J}^{-1}
& = \sum_{a}d_{a}(\bm{\mathrm{k}}) \Bigr(\mathcal{J}\Gamma^{a}\mathcal{J}^{-1}\Bigr)
= h(\bm{\mathrm{k}}).
\end{align}

The negative-energy doublet defines the complex rank-two spectral projector 
\begin{align}
& P (\bm{\mathrm{k}}) = \frac{1}{2}(\bm{1}_{4} - \hat{d}_{a}(\bm{\mathrm{k}})\Gamma^{a}), \quad \hat{d}_{a}(\bm{\mathrm{k}}) = \frac{d_{a}(\bm{\mathrm{k}})}{|d(\bm{\mathrm{k}})|}.
\end{align}
Since $\mathcal{J}P(\bm{\mathrm{k}})\mathcal{J}^{-1} = P(\bm{\mathrm{k}})$, its image is a quaternionic line, and hence 
\begin{align}
& P_{\mathbb{H}}: T^{4} \rightarrow \mathbb{H}P^{1}.
\end{align}
Using 
\begin{align}
& \partial_{\mu}P = -\frac{1}{2}(\partial_{\mu}\hat{d}_{a})\Gamma^{a}, \quad
\mathrm{Tr}(\Gamma^{a}\Gamma^{b}) = 4\delta^{ab},
\end{align}
we obtain the quantum metric
\begin{align}
g_{\mu\nu} 
& = \mathrm{Re}\mathrm{Tr}(P\partial_{\mu}P \partial_{\nu}P) \nonumber \\
& = \frac{1}{2}\mathrm{Tr}(\partial_{\mu}P \partial_{\nu}P)
= \frac{1}{2}(\partial_{\mu}\hat{d}_{a})(\partial_{\nu}\hat{d}_{a}).
\end{align}
Let us define the $5\times5$ matrix
\begin{align}
& \mathcal{G}(\bm{\mathrm{k}}) = (\hat{d}(\bm{\mathrm{k}}), \partial_{x}\hat{d}(\bm{\mathrm{k}}), \partial_{y}\hat{d}(\bm{\mathrm{k}}), \partial_{z}\hat{d}(\bm{\mathrm{k}}), \partial_{w}\hat{d}(\bm{\mathrm{k}})).
\end{align}
Since $\hat{d}\cdot\hat{d}=1$, we have $\hat{d} \cdot (\partial_{\mu}\hat{d})=0$. Using these relations, we obtain \cite{Ding2024NonAbelianQGT, Zhang2022RevealingChernNumber}
\begin{align} \label{determinant of g in T4}
(\det \mathcal{G})^{2} 
& = \det(\mathcal{G}^{T}\mathcal{G}) 
= \det 
\begin{pmatrix}
\hat{d} \cdot \hat{d} & \hat{d} \cdot (\partial_{\nu}\hat{d}) \\
(\partial_{\mu}\hat{d}) \cdot \hat{d} & (\partial_{\mu}\hat{d}) \cdot (\partial_{\nu}\hat{d})
\end{pmatrix} \nonumber \\
& = \det
\begin{pmatrix}
1 & 0 \\
0 & (\partial_{\mu}\hat{d}) \cdot (\partial_{\nu}\hat{d})
\end{pmatrix}
= \det
\begin{pmatrix}
1 & 0 \\
0 & 2g_{\mu\nu}
\end{pmatrix} \nonumber \\
& = \det (2g) = 16\det g.
\end{align}

The second Chern density is given by
\begin{align} \label{explicit FwedgeF}
\mathrm{Tr} (F\wedge F)
& = \mathrm{Tr} (P (\mathrm{d}P)^{4}) \nonumber  \\
& = \frac{1}{8} \epsilon^{abcde} \hat{d}_{a}\mathrm{d}\hat{d}_{b} \wedge \mathrm{d}\hat{d}_{c} \wedge \mathrm{d}\hat{d}_{d} \wedge \mathrm{d}\hat{d}_{e} \nonumber \\
& = \frac{4!}{8} (\det \mathcal{G})\mathrm{d}^{4}k
= 3 (\det \mathcal{G})\mathrm{d}^{4}k,
\end{align}
where $\mathrm{d}^{4}k = \mathrm{d}k_{x}\wedge\mathrm{d}k_{y}\wedge\mathrm{d}k_{z}\wedge\mathrm{d}k_{w}$. Therefore, for all $\bm{\mathrm{k}} \in T^{4}$, the quaternionic Wirtinger inequality saturates:
\begin{align}
& \frac{1}{12}|[\mathrm{Tr}(F \wedge F)]_{1234}| = \sqrt{\det g} = \frac{1}{4}|\det \mathcal{G}(\bm{\mathrm{k}})|,
\end{align}
where 
\begin{align} \label{det mathcal g}
\det\mathcal{G}(\bm{\mathrm{k}}) 
= \frac{1}{|d(\bm{\mathrm{k}})|^{5}} &\Biggr[ \Bigr( m
+c\sum_{\mu}\cos k_{\mu} \Bigr) \prod_{\nu}\cos k_{\nu} \nonumber \\
& + c\sum_{\mu}\sin^{2}k_{\mu} \prod_{\nu \ne \mu} \cos k_{\nu}\Biggr].
\end{align}
The sign change can be seen explicitly near $\bm{\mathrm{k}}_{*} = (\pi/2,\pi/2,0,0)$. For
\begin{align}
& \bm{\mathrm{k}}_{\pm} = \left(\frac{\pi}{2}\pm\epsilon,\frac{\pi}{2},0,0\right), \quad 0<\epsilon<\frac{\pi}{2},
\end{align}
Eq. (\ref{det mathcal g}) gives
\begin{align} \label{det g kpm}
& \det \mathcal{G}(\bm{\mathrm{k}}_{\pm}) = \mp \frac{c\sin \epsilon}{|d(\bm{\mathrm{k}}_{\pm})|^{5}}.
\end{align}
Hence, $\det \mathcal{G}$ vanishes at $\bm{\mathrm{k}}_{*}$ and changes sign across it. Equation (\ref{determinant of g in T4}) then gives $\det g(\bm{\mathrm{k}}_{*})=0$, while Eq. (\ref{explicit FwedgeF}) shows that the second Chern density also vanishes and changes sign there.

\section{Conventions for Quaternionic K\"ahler Geometry in Four Dimensions} \label{appendix F}
In this Appendix, we compare the geometric structures described as quaternionic K\"ahler in Refs.~\cite{Zhao2026SecondChernBoundsNonAbelian,Yang2026QKgeometryTRS} with the terminology and definitions adopted in the present work. 

Ref. \cite{Zhao2026SecondChernBoundsNonAbelian} provides an independent and complementary derivation of an equivalent determinant bound through the Hodge decomposition of the $\mathrm{SU}(2)$ Berry curvature. At points where the induced metric is nondegenerate, its saturation conditions consist of the Hodge self-duality of the Berry curvature and the condition that the induced tangent space endomorphisms form a local admissible frame of a metric-compatible quaternionic structure. The resulting structure is described therein as quaternionic K\"ahler. In the terminology of the present work, the resulting metric-compatible quaternionic structure is quaternionic Hermitian in the sense of Definition \ref{definition 4}. The Hodge self-duality condition in Ref. \cite{Zhao2026SecondChernBoundsNonAbelian} concerns the $\mathrm{SU}(2)$ Berry curvature rather than the self-duality of the Riemannian Weyl tensor; hence, our curvature-based four-dimensional convention of Definition \ref{definition 7} additionally requires the induced metric to be Einstein with self-dual Weyl curvature. To clarify the convention underlying this distinction, we next turn to Ref. \cite{Yang2026QKgeometryTRS}, which explicitly distinguishes operational and standard definitions of quaternionic K\"ahler geometry. 

Let $(X,g)$ be a Riemannian manifold of real dimension $4n$. Ref. \cite{Yang2026QKgeometryTRS} operationally defines $(X,g)$ to be quaternionic K\"ahler if it admits a rank-three subbundle
\begin{align}
& \mathcal{Q} \subset \mathrm{End}(TX)
\end{align}
that is locally spanned on an open cover $\{U_{\alpha}\}$ by three $g$-compatible endomorphisms satisfying 
\begin{align}
& I^{i}_{\alpha}I^{j}_{\alpha} = -\delta^{ij}\mathrm{id}_{TX} + \epsilon^{ijk}I^{k}_{\alpha}, \
 \omega^{i}_{\alpha}(v,w) = g(v,I^{i}_{\alpha}w),
\end{align}
for $x \in U_{\alpha}$, $v,w\in T_{x}X$, and $i,j=1,2,3$. The convention for $\omega^{i}_{\alpha}$ in Ref. \cite{Yang2026QKgeometryTRS} differs by an overall sign from Eq. (\ref{relation btw w and g}), without affecting the underlying bundle $\mathcal{Q}$. On every nonempty overlap $U_{\alpha}\cap U_{\beta}$, the local triples are related by an $\mathrm{SO}(3)$ rotation,
\begin{align}
& I^{i}_{\alpha} = (R_{\alpha\beta})^{i}_{\ j} I^{j}_{\beta}, \quad 
R_{\alpha\beta}: U_{\alpha}\cap U_{\beta} \rightarrow \mathrm{SO}(3),
\end{align}
ensuring that the local spans $\mathcal{Q}|_{U_{\alpha}} = \mathrm{span}\{ I_{\alpha}^{1}, I_{\alpha}^{2}, I_{\alpha}^{3} \}$ agree on overlaps. No condition involving the Levi-Civita derivative of $\mathcal{Q}$ is imposed in this operational definition. 
Once such a bundle $\mathcal{Q}$ is specified, the resulting structure $(X,g,\mathcal{Q})$ is quaternionic Hermitian in the sense of Definition \ref{definition 4}.

Ref. \cite{Yang2026QKgeometryTRS} further contrasts its operational definition with the standard definition, which additionally requires $\mathcal{Q}$ to be preserved by the Levi-Civita connection:
\begin{align}
& \nabla I^{i}_{\alpha} = (\theta_{\alpha})^{i}_{\ j}\otimes I^{j}_{\alpha}, 
\end{align}
where $\theta_{\alpha}$ is a skew-symmetric matrix of 1-forms on $U_{\alpha}$. Equivalently, it is characterized by the holonomy reduction
\begin{align} \label{holonomy condition of qk}
& \mathrm{Hol}(g) \subset \mathrm{Sp}(n)\cdot \mathrm{Sp}(1).
\end{align}
In the terminology adopted in the present work, the additional condition upgrades a quaternionic Hermitian structure to a quaternionic K\"ahler structure of Definition \ref{definition 6} for $n\ge2$. Its four-dimensional specialization, $n=1$, will be discussed below.

For $n=1$, the relation between the operational and standard definitions requires some care. Once an almost quaternionic structure $\mathcal{Q}$ is fixed, the operational definition amounts precisely to the quaternionic Hermitian compatibility condition of Definition \ref{definition 4}. The additional Levi-Civita preservation condition required by the standard definition, however, becomes automatic in real dimension four. Indeed,
\begin{align} 
& \mathrm{Hol}(g) \subset \mathrm{Sp}(1)\cdot\mathrm{Sp}(1) \cong \mathrm{SO}(4),
\end{align}
so Eq. (\ref{holonomy condition of qk}) imposes no further restriction on a four-dimensional quaternionic Hermitian manifold. Thus, the standard definition does not strengthen the operational one by an additional differential condition in four dimensions.

To retain a nontrivial four-dimensional distinction between quaternionic Hermitian and quaternionic K\"ahler geometry, we follow the four-dimensional curvature criterion discussed by Swann \cite{Swann1990QK} and reserve the term quaternionic K\"ahler in real dimension four for the curvature-based convention of Definition \ref{definition 7}. Accordingly, a four-dimensional quaternionic Hermitian manifold $(X,g,\mathcal{Q})$ is quaternionic K\"ahler in our terminology only when $g$ is additionally Einstein and self-dual with respect to the orientation determined by $\mathcal{Q}$. With this dictionary, the quaternionic K\"ahler band geometry in the operational terminology of Ref. \cite{Yang2026QKgeometryTRS} corresponds to quaternionic Hermitian band geometry in the present work; our four-dimensional quaternionic K\"ahler terminology imposes the additional Einstein and self-duality conditions.

This difference in four-dimensional terminology does not affect our use of Proposition S7 of Ref. \cite{Yang2026QKgeometryTRS} in the proof of Theorem \ref{theorem 4}.
Its rigidity conclusion relies only on the nondegeneracy of the induced metric, saturation of the quaternionic Wirtinger inequality, and the resulting invariance of $\mathrm{d}P_{\mathbb{H}}(TU)$ under the local quaternionic structures $\mathcal{Q}_{\mathrm{FS}}$ of $\mathbb{H}P^{N-1}$. It requires neither the Einstein condition nor the self-duality of the Weyl curvature of the induced four-dimensional metric. Hence, although the resulting structure is called quaternionic K\"ahler in the operational terminology of Ref. \cite{Yang2026QKgeometryTRS} and quaternionic Hermitian in the present work, Proposition S7 applies without modification.

%Finally, some authors, including Swann \cite{Swann1990QK}, further reserve the term quaternionic K\"ahler for the nonzero scalar curvature, separating it from scalar-flat and hyperK\"ahler cases. This refinement does not affect the present $S^{4}$ result, since the induced metric has positive scalar curvature, but may become relevant in future extensions.

%\bibliography{bibliography}
\bibliography{refs}

\end{document}